\documentclass[12pt]{extarticle}
\usepackage{hyperref}
\usepackage[utf8]{inputenc}
\usepackage[english]{babel}
\usepackage{amsmath}
\usepackage{amssymb}
\usepackage{graphicx}
\usepackage{caption}
\usepackage{verbatim}
\usepackage{comment}
\usepackage{tabularx,booktabs}
\usepackage{easytable}
\usepackage{bm}
\usepackage{subfigure}
\usepackage{amstext, mathrsfs, textcomp}
\usepackage{multirow}
\usepackage{xcolor}
\usepackage{enumerate}
\usepackage{tabularx}
\usepackage{textcomp}
\usepackage{siunitx}
\usepackage[normalem]{ulem}

\usepackage{blindtext}
\usepackage{comment}
\usepackage{multirow}
\usepackage{booktabs}
 \usepackage{multirow}
\usepackage{graphicx}
\usepackage{xcolor}
\usepackage{bm}
\usepackage{braket}
\usepackage{multirow,tabularx}
\usepackage{tabu}
\usepackage{longtable}

\newcommand{\red}[1]{{\color{red}#1}}

\newcommand{\ve}[1]{\mathbf{#1}}
\renewcommand{\t}[1]{\text{#1}}

\newcommand{\om}{\omega}

\newcommand{\veps}{\varepsilon}

\let\vec=\mathbf

\title{All-dielectric thermonanophotonics}

\author{George P. Zograf,~$^1$ Mihail I. Petrov,~$^1$ \\  Sergey V. Makarov,~$^1$ and  Yuri S. Kivshar~$^{1,2,*}$}

\begin{document}
\maketitle
$^1$~Department of Physics, ITMO University, Saint-Petersburg, 197101, Russia
\\
$^2$~Nonlinear Physics Centre, Australian National University, Canberra ACT 2601, Australia
\\
$^*$yuri.kivshar@anu.edu.au 



\section*{Abstract}
Nanophotonics is an important branch of modern optics dealing with light-matter interaction at the nanoscale.  Nanoparticles can exhibit enhanced light absorption under illumination by light, and they become nanoscale sources of heat that can be precisely controlled and manipulated. For metal nanoparticles, such effects have been studied in the framework of {\it thermoplasmonics} which, similar to plasmonics itself, has a number of limitations. Recently emerged {\it all-dielectric resonant nanophotonics} is associated with optically-induced electric and magnetic Mie resonances, and this field is developing very rapidly in the last decade. As a result, thermoplasmonics is being replaced by {\it all-dielectric thermonanophotonics} with many important applications such as photothermal cancer therapy, drug and gene delivery, nanochemistry, and photothermal imaging. This review paper aims to introduce this new field of non-plasmonic nanophotonics and discuss associated thermally-induced processes at the nanoscale.

\newpage
\tableofcontents
\newpage

\section{Introduction}

Nanophotonics deals with optically resonant nanostructures, and it provides useful tools to control light at the nanoscale. For many years, nanophotonics was associated with metallic structures and their ability to support surface plasmon polaritons, or plasmons, which are hybrid modes created by coupling of electromagnetic waves to free electrons in metals. Metallic nanostructures supporting plasmons are well-known for their ability to achieve extreme light localization~\cite{baumberg2019extreme}, enhance emission~\cite{giannini2011plasmonic}, manipulate scattering of light~\cite{schuller2010plasmonics}, demonstrate nonlinear effects~\cite{kauranen2012nonlinear}, and also improve efficiencies of solar cells~\cite{atwater2010plasmonics}. 

Energy of incident light can be localized efficiently in resonant nanostructures, and this results in heating of the nanostructures~\cite{baffou2020applications}.  A branch of plasmonics that studies thermal effects was shaped as an independent field named {\it thermoplasmonics}. It describes many novel effects and has unique applications such as photothermal therapy~\cite{huang2008plasmonic}, catalysis~\cite{aslam2018catalytic}, reshaping of optical responses~\cite{gonzalez2017femtosecond}, and many others.

Conversion of intense energy of light into thermal heat in plasmonic nanostructures is well described for the steady-state regime which represents continuous-wave illumination~\cite{baffou2010nanoscale, govorov2007generating}, and also for the case of pulsed laser heating~\cite{baffou2011femtosecond,tribelsky2011laser}. One of the main approaches for highly efficient optical heating of plasmonic structures is the excitation of surface plasmon resonances manifested by a sufficient increase of optical absorption. In the case of small (a $\ll$ $\lambda$) nanoparticles, the surface plasmon resonance can be  tuned gradually via prolongation of the nanoparticle (NP) along one axis covering the whole visible and near-IR region. Owing to very small sizes, plasmonic nanoparticles can find a number of applications.

Optical heating of plasmonic nanostructures is widely used in chemistry for photothermal catalysis~\cite{baffou2014nanoplasmonics}, water heating~\cite{ishii2016titanium,ni2015volumetric, tao2018solar}, thermophotovoltaics~\cite{zhou2016solar, omair2019ultraefficient}, data recording~\cite{zijlstra2009five}, surface coloring~\cite{kristensen2016plasmonic}, and many other fields~\cite{chen2021plasmonic}. 

One of the most important applications of thermoplasmonics is photothermal therapy, which uses local overheating for triggering physiological process with proteins starting from stimulation of their diffusion through cell membrane up to their damage and denaturation~\cite{li2015optical}. Such local overheating of plasmonic nanoparticles was used for cancer antitumor therapy by means of near-infrared light exposure of the desired region with delivered plasmonic NPs~\cite{pitsillides2003selective,hirsch2003nanoshell}. Recent trends lead to combination of thermoplasmonics with various approaches in cancer therapy~\cite{beik2019gold}. Finally, plasmonic nanostructures are efficient in photothermal sensing of notorious COVID-19~\cite{qiu2020dual}. However, these are not the only applications of thermoplasmonics in therapy. For instance, local plasmonic overheating is used for bacterial sterilization~\cite{pihl2017bacterial}, skincare of acne~\cite{paithankar2015acne}, hair removal~\cite{harris2016hair} or even retinal treatment~\cite{wilson2018vivo}.

On the other hand, biomedical applications require a control of local temperatures in order to avoid unnecessary hyperthermia. In this regard, plasmonic NPs are usually considered as objects with thermally inactive inherent optical response, such as Raman scattering of photoluminescence (PL). Therefore, additional thermally-responsive coatings or substances were employed to estimate the local temperature around various metal nanostructures~\cite{debasu2013all, baffou2013photoinduced}. However, recently a method of thermally-sensitive anti-Stokes PL in gold nanorods which can be used to measure the local temperature was proposed~\cite{carattino2018gold}. On this way, further progress on dramatic enhancement of PL efficiency form metal nanoparticle should be done to make it suitable for above mentioned thermoplasmonics applications.

\begin{figure}
\centering
  \includegraphics[width=.85\textwidth]{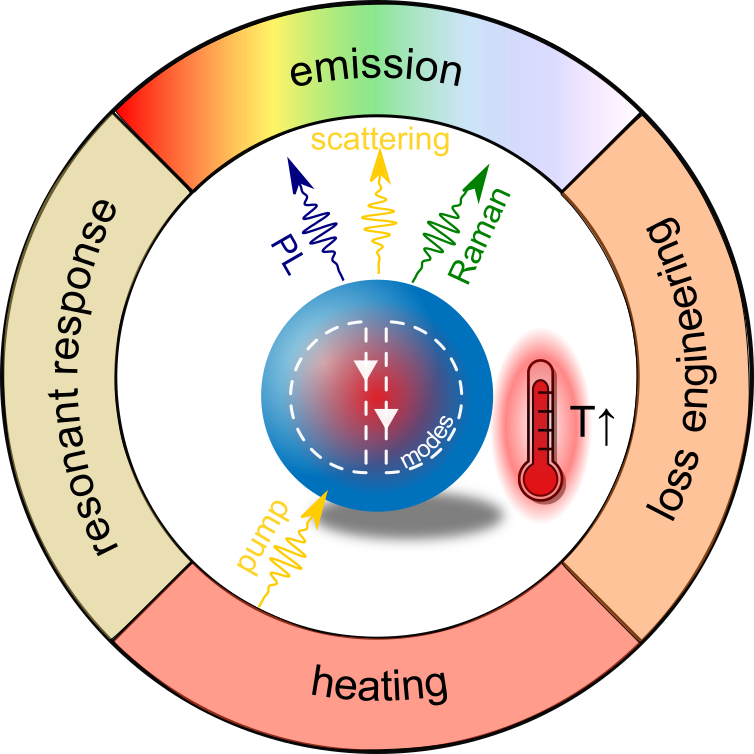}
  \caption{%
The key concepts underpinning the physics of optical heating of all-dielectric nanoparticles, with the most important effects for optimization and applications.
}
  \label{fig:concept}
\end{figure}

{\it All-dielectric nanophotonics} has been rapidly developing for the last several years~\cite{kuznetsov2016optically, staude2017metamaterial}. Pursuing new high-efficient compact optical systems, the research has been mainly driven by resonant high-refractive index dielectric and semiconductor nanostructures, which are relieved of parasitic Ohmic losses inherent by plasmonics. On this way, outstanding progress has been achieved recently in ultracompact light-emitting systems~\cite{staude2015shaping, tiguntseva2018light, tiguntseva2018tunable, vaskin2018directional, liu2018light,vaskin2019light, staude2019all, bucher2019tailoring, ha2018directional, ma2019applications, paniagua2019active, hoang2020collective, wu2020room, tiguntseva2020room}, optical sensing~\cite{walker2003quantum,ha2016laser,balkanski1983anharmonic, rodriguez2014silicon, zograf2017resonant, bontempi2017highly, yavas2017chip, tittl2018imaging, mitsai2018chemically,yavas2019unravelling, yesilkoy2019ultrasensitive, garcia2019enhanced}, nonlinear nanophotonics~\cite{shcherbakov2014enhanced, smirnova2016multipolar, li2017nonlinear, grinblat2017efficient,makarov2017efficient, liu2018all, kruk2019nonlinear, xu2019forward, sain2019nonlinear, koshelev2020subwavelength}, and photovoltaics~\cite{furasova2018resonant}. Despite the fact that losses in all-dielectric materials can be negligible in certain spectral ranges, their spectral dispersion is extremely strong, especially in semiconductor materials in the vicinity of direct (or indirect) transition threshold. Being one of the major problems in solid-state optoelectronics, in nanophotonics the optical losses opened a route for efficient optical heating of subwavelength all-dielectric nanostructures. Till recently, the area of thermal nano-optics was fully associated with nanoplasmonics. Indeed, due to the ability for enormous light localization near the metal surface, plasmonics structures suffered from the parasitic side effect of overheating in applications such as sensing, imaging, spectroscopy and photovoltaics. However, in the late 1990s and early 2000s~\cite{boyer2002photothermal,huttmann1999possibility} there have been found promising biomedical applications for precise controllable light-induced protein unfolding and temperature probing and imaging for hyper-thermal therapies. Since then, a huge field of nano-optics studying nanoscale thermal effects in plasmonics structures has emerged~\cite{baffou2020applications,baffou2013thermo}.

Some of the applications of thermoplasmonics are already being used in industry, whereas the field of all-dielectric thermonanophotonics based resonant dielectric nanoparticles is new, but it may provide additional degrees of freedom due to a number of features, including 

\begin{itemize}

\item
non-metallic optical materials with a broad range of functionalities;  

\item
low optical losses of dielectric materials precisely tunable from zero to extremely high values; 

\item
strong optical nonlinearity and thermorefractive properties of non-metallic materials.

\end{itemize}

In this paper, we review the fundamentals of optical heating of subwavelength resonant non-metallic nanostructures and provide a summary of the recent advances in application of dielectric structures for nanoscale heating and thermometry. 
By replacing partially plasmonic structures and thermoplasmonic effects, all-dielectric thermonanophotonics provides novel degrees of freedom for tuning the optical responses by employing Mie resonances, and it offers \textit{in situ} temperature detection by means of photoluminescence or Raman scattering. Different strategies to achieve the best performance are provided with the focus on a balance between radiative and non-radiative losses in resonant nanoparticles.

\section{Fundamentals of optical heating at the nanoscale}



Optical heating process represents a cascade of various processes starting from photon absorption and finishing when system relaxes in its initial state after complete cooling.

\subsection{Ultrafast optical heating: Two-temperature regime}

In metals, light is almost exclusively absorbed by free electron transitions within the conduction band. The electron system is thermalized, typically, during 10~fs to 1 ps. Thermalization between the electron subsystem and the lattice is much slower, typically of the order of 1-100 ps, depending on the strength
of electron-phonon coupling. Thus, femtosecond laser excitation generates a hot electron gas, which than heats the ion (or lattice) sub-system. The coupled nonlinear equations for electrons and lattice can be written, in a more general form, as~\cite{anisimov1974electron}

\begin{equation}\label{Eq:TTM}
  \begin{cases} C_e \partial T_e / \partial t= \nabla (\kappa_e \nabla T_e)-\gamma_{ei}(T_e-T_i)+S \\
    C_i \partial T_i / \partial t= \nabla (\kappa_i \nabla T_i)+\gamma_{ei}(T_e-T_i)
      \end{cases}
      \end{equation}
where $T_e$ and $T_i$ are the electron and lattice temperatures, respectively. $C_i$ and $\kappa_i$ denote the specific heat capacity and the thermal conductivity of the subsystem $j (j = e, i)$ respectively. The laser pulse is modeled as a source term, $S$, particular for each material. 
Laser light is absorbed by the electronic subsystem which transfers the absorbed energy to the lattice through electron-phonon
collisions represented by the coupling factor $\gamma_{ei}$.

In dielectrics and semiconductors, ultrafast thermal modeling should involve the adequate
mechanisms of the photoexcited electron generation and recombination. The electron density in the conduction band $N_e$ should be calculated
considering absorption term, carrier diffusion,
impact ionisation and Auger recombination by solving the following partial differential equation:
\begin{equation}\label{Eq:carriers}
         \partial N_e / \partial t = \nabla(k_B T_e \mu_e \nabla N_e) + G_e-R_e
\end{equation}
the term $G_e$ describes the electron generation rate (can be nonlinear), and $R_e$ corresponds to relaxation processes such as Shockley-Read-Hall effect, radiative recombination and Auger recombination. The first term on the right-hand side of Eq.~\ref{Eq:carriers} describes the carrier transport due to diffusion where $k_B$ is the Boltzmann constant and $\mu_e$ is the electron mobility in the conduction band. Also, the Eq.~\ref{Eq:carriers} determines such a basic property as electron specific heat capacity:
\begin{equation}\label{Eq:capacity}
C_e = \frac{3}{2}k_B N_e
\end{equation}
and electron-phonon coupling factor~\cite{yoffa1981screening,sjodin1998ultrafast}:
\begin{equation}\label{Eq:gamma}
\gamma_{ei} = \frac{C_e}{\tau_{\gamma}} = \frac{C_e}{\tau_0 [1+(N_e/N_{th})^2]}
\end{equation}
where $\tau_0$ is the hot carrier relaxation time and $N_{th}$ is the critical carrier density for screening of electron-phonon interaction. It means that the screening increases total effective time of electron-phonon relaxation in semiconductors $\tau_{\gamma}$. For example, in such semiconductors as Si and GaAs, $\tau_{\gamma}$ is typically between 0.1~ps and 2~ps depending on photogenerated free electron density~\cite{sjodin1998ultrafast, bernardi2015ab}.

\begin{figure}
\centering
  \includegraphics[width=.89\textwidth]{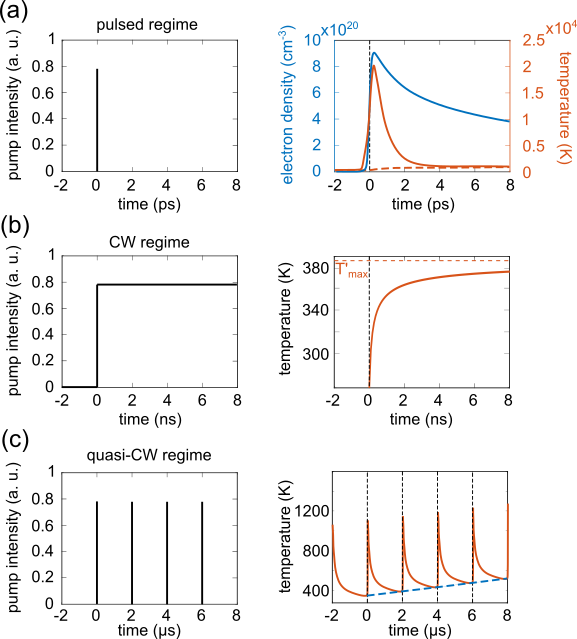}
  \caption{%
\textbf{Regimes of optical heating.} (a) Calculated carrier number density, carrier temperature, and lattice temperature for bulk silicon surface irradiated by 350-fs laser pulse with fluence $\sim$0.5~J/cm2 schematically shown in upper inset.~\cite{taylor2018integrating} (b) Calculated solution of Eqs.~\ref{Eq:CWheat},\ref{Eq:Tmax} for the case of silicon surface irradiated by 515-nm CW laser with intensity 50~mW/$\mu$m$^2$ and Gaussian beam radius 0.7~$\mu$m. (c) Calculated surface maximum temperature of bulk silicon versus time upon irradiation by a train of 350-fs laser pulses with fluence $\sim$0.5~J/cm2 coming with repetition rate 500~kHz schematically shown in upper inset. Red dash line indicates growth of background temperature~\cite{taylor2018integrating}.}
  \label{fig:Time}
\end{figure}

Figure~\ref{fig:Time}a shows solution of Eqs.\ref{Eq:TTM}-\ref{Eq:gamma} for Si surface irradiated by an intense fs-laser pulse, where typical scenario of ultrafast electron subsystem heating and subsequent energy transfer to lattice is presented. Remarkably, temperatures of electron and lattice become equal at the picosecond scale. 
However, further cooling of the heated bulk material is strongly dependent on its thermal conductivity.

Cooling rate $K_c$ of semiconductor-based nanoparticles supporting optical resonances is around 0.1--1~K/ps~\cite{makarov2018resonant, larin2020plasmonic}, being strongly dependent on thermal conductivity of the substrate or hosting medium. Cooling rate of plasmonic nanoparticles under pulsed illumination is discussed in detail elsewhere~\cite{baffou2013thermo}. As a result, despite the ultrafast character of optical heating upon fs-laser pulses, the effect of temperature localization at nanoscale can be diminished due to too high repetition rate of the coming laser pulses.

\subsection{Slow optical heating: One-temperature regime}

In many cases and practical applications, quasi-CW and CW optical heating are more preferable or realistic regimes of optical heating. In this regard, it is crucial to understand what are the stages in the time-evolution of gradual temperate growth and what are the main limitations in this process. 

For solving a problem of optical heating lets start with general equation for the lattice temperature only:

\begin{equation}\label{Eq:HeatT}
 \rho c_{p} \partial T / \partial t=\nabla(\kappa \nabla T)+(\partial Q / \partial t),  
\end{equation}
where $\rho$ is material density, $c_p$ is heat capacity, $\kappa$ is thermal conductivity, $T$ is temperature and $Q$ is the amount of heat received per unit volume of material. Temperature dependence of $\rho$, $c_p$, $\kappa$, and $(\partial Q / \partial t)$ is negligible in many cases. Thus, Eq.\ref{Eq:HeatT} can be rewritten to the form:
\begin{equation}\label{Eq:HeatT2}
    \partial T / \partial t=\chi \Delta T+\left(\rho c_{p}\right)^{-1}(\partial Q / \partial t),
\end{equation}
where $\chi=\kappa / \rho c_{p}$ is thermal diffusivity.
Laser beam propagation along the axis z and fall on xy plane of material create volume heat source:
\begin{equation}
   \partial Q / \partial t=\alpha I(r, t),
\end{equation}
where $\alpha$ is absorption coefficient and $I(r, t)$ is distribution of light intensity (in region z>0). One of the most widely used case is Gaussian beam, which can be written as:
\begin{equation}\label{gauss}
    I(r, t)=(1-R) I_{0} \exp (-\alpha z) \exp \left[-\left(x^{2}+y^{2}\right) / a^{2}\right] f\left(t / \tau_{p}\right),
\end{equation}
where $I_0$ is the intensity of radiation incident on the absorbing medium from the outside along $z$ direction, $R$ is optical reflection coefficient and $a$ is radius of Gaussian beam. Function $f \left(t / \tau_{p} \right)$ is a general form of temporal shape of laser pulse with duration $\tau_p$.

If Eq.~\ref{Eq:HeatT} is linear, it is also valid for the increase of the sample temperature $T'=T-T_0$, where $T_0$ is the sample temperature before the optical irradiation. Also, in many cases, the process of heating a material can be considered in the absence of heat exchange with the environment using the following boundary condition:
\begin{equation}\label{Eq:boundCond}
  \left.\chi \frac{\partial T^{\prime}}{\partial z}\right|_{z=0}=0
\end{equation}
Method of Green function is convenient for solving Eq.\ref{Eq:HeatT}. Indeed, let's assume that the Green function $G\left(r-r^{\prime}, t-t^{\prime}\right)$ is the solution for a heat source perfectly localized in $r=0$ and in the form of instant pulse (i.e. defined by delta-functions $\delta$):
\begin{equation}
    \frac{1}{\rho c_{p}}\left(\frac{\partial Q}{\partial t}\right) \rightarrow A \delta\left(\boldsymbol{r}-\boldsymbol{r}^{\prime}\right) \delta\left(t-t^{\prime}\right)
\end{equation}
When Eq.\ref{Eq:HeatT2} with boundary conditions Eq.\ref{Eq:boundCond} is linear, the solution of the heat problem with arbitrary source has the form:

\begin{equation}
  T'(r, t)=\frac{1}{\rho c_{p}} \int_{-\infty}^{t} d t^{\prime} f\left[\frac{\partial Q}{\partial t}\left(r^{\prime}, t^{\prime}\right)\right] G(r-r, t-t) d^{3} r^{\prime}  
\end{equation}

For a point source, Eq.\ref{Eq:HeatT2} can be rewritten as:

\begin{equation}
    \partial T^{\prime} / \partial t=\chi \Delta T^{\prime}+A \delta(r) \delta(t)
\end{equation}

Solving this equation with taking into account the boundary conditions gives the following expression:

\begin{equation}
    T^{\prime}(r, t)=\frac{A}{(4 \pi \chi t)^{3 / 2}} \exp \left(-\frac{r^{2}}{4 \chi t}\right)
\end{equation}

According to this expression, after instant point-like heat source switching on, the temperature rise in the heated area has temporal dependence like $T' \sim t^{-3/2}$, while characteristic size of the heated zone grow with time as $\sim (\chi t)^{1/2}$. 

General solution of Eq.\ref{Eq:HeatT2} for the continuous wave laser excitation with Gaussian intensity profile (see Eq.\ref{gauss}) can be written as:
\begin{equation}\label{solution}
    \begin{array}{l}{\frac{\partial T'}{\partial t}=\frac{\alpha(1-R) I_{0}}{2 \rho_{0} c_{p}} \frac{1}{1+4 \chi t / a^{2}} \exp \left(-\frac{x^{2}+y^{2}}{a^{2}+4 \chi t}\right)} { \exp \left(\alpha^{2} \chi t\right)\left\{\exp (\alpha z) \operatorname{erfc}\left[\alpha(\chi t)^{1 / 2}+\frac{z}{(4 \chi t)^{1 / 2}}\right]+\right.} \\ {\left.+\exp (-\alpha z) \operatorname{erfc}\left[\alpha(\chi t)^{1 / 2}-\frac{z}{(4 \chi t)^{1 / 2}}\right]\right\} \theta(t)}\end{array},
\end{equation}
where $\theta(t)$ is Heaviside function and $ \operatorname{erfc} x=\frac{2}{\sqrt{\pi}} \int_{x}^{\infty} e^{-t^{2}} \mathrm{d} t$ is complementary error function. 

The maximum rate of temperature increase is on the surface of the irradiated substrate ($z$=0) and on the axis of the laser beam incidence (x=y=0):
\begin{equation}\label{Eq:lowtime}
    \frac{\partial T'}{\partial t} (r=0, t)=\frac{\alpha(1-R) I_{0}}{\rho_{0} c_{p}} \frac{\exp \left(\alpha^{2} \chi t\right) \operatorname{erfc}\left[\alpha(\chi t)^{1 / 2}\right]}{1+4 \chi t / a^{2}}
\end{equation}
At the initial stage ($t \leqslant \min \left\{a^{2} / \chi, 1 / \alpha^{2} \chi\right\}$) thermal conductivity does not affect the heating rate, which is almost constant ($ \partial T' / \partial t=\alpha(1-R) I_{0} / \rho_{0} c_{p}$ from Eq.\ref{Eq:lowtime} at $t$=0). The physical meaning of the quantity $a^2/\chi$ is the characteristic time during which the temperature transfer occurs up to the distance corresponding to lateral size of the laser beam $a$.

For the case of highly absorbing materials like metals or semiconductors at wavelengths in the interband absorption range, penetration depth of light in the material ($d=\alpha^{-1}$ absorption length) manifests itself in the sample heating after some characteristic time $d^2/\chi=1/\alpha^2 \chi$. Thus, at the times $t \leqslant \min \left\{a^{2} / \chi, 1 / \alpha^{2} \chi\right\}$ the maximum temperature of the substrate increases over time according to the linear law: $T' \sim t$ as follows from Eq.\ref{solution}.

Let us assume that the transverse size of the laser beam is significantly greater than the absorption length ($a \gg d$). Then, with time increasing the heat transfer into the medium "switch on" of the first time is included which reduces the rate of heating. For the condition $1 / \alpha^{2} \chi \leqslant t \leqslant a^{2} / \chi$ Eq.\ref{Eq:lowtime} takes the form, where the temperature increment grows sub-linearly ($T' \sim t^{1/2}$). During the time $t\geq a^2 /\chi$ after the laser exposure, thermal conductivity in the direction along the surface begins to influence the temperature increase. Heating rate decreases rapidly. At $t\geq a^2/\chi$ the solution for temperature $t \gg a^2/\chi $  is established by equation:

\begin{equation}\label{Eq:CWheat}
    T'\approx T'_{\max }-\frac{(1-R) I_{0} \pi a^{2}}{2 \rho_{0} c_{p} \pi \chi(4 \pi \chi t)^{1 / 2}}
\end{equation}

The maximum temperature $T'_{max}$ are defined as integral Eq.\ref{Eq:lowtime}: $T'_{max}=T'(r=0,t=\infty$). In case $\alpha a \gg 1$, it can be estimated as: 

\begin{equation}\label{Eq:Tmax}
    T'_{\max } \sim \frac{(1-R) I_{0} a}{\rho_{0} c_{p} \chi}=\frac{(1-R) \mathfrak{P} }{\kappa a},
\end{equation}
where $\mathfrak{P}=a^2 I_0$ is total power of Gaussian beam with $a$ radius and $I_0$ intensity.

For the opposite case, when ($d \gg a$) slowing of the heating rate starts at $t \geq a^2/\chi$ and yielding the logarithmic trend $T'\sim\ln(t)$, when the relatively slow temperature growth caused by more efficient lateral heat flow to the bulk. Further heating in the regime of $t\gg 1/\alpha^2\chi$ leads to saturation of temperature following the law Eq.\ref{Eq:CWheat} at $ln(1/\alpha a)\gg 1$ with maximum temperature:

\begin{equation}\label{Eq:Tmax2}
    T'_{\max } \sim \frac{(1-R) \mathfrak{P} }{\kappa}\alpha \rm{ln}\frac{1}{\alpha a},
\end{equation}

Solution of Eq.~\ref{Eq:CWheat} for Si substrate and 515-nm CW laser with intensity 50~mW/$\mu$m$^2$ and Gaussian beam radius 0.7~$\mu$m is given in Fig.~\ref{fig:Time}b. The maximum temperature increase from the 273~K in this case is around 120 K, which is at least one order of magnitude less than that for the pulsed regime (see Fig.~\ref{fig:Time}a). 

In many cases, laser irradiation of targets is carried out by train of short pulses. This regime is a mixture of pulsed optical heating and slow continuous growth of temperate, which is usually called as quasi continuous wave (quasi-CW) regime. In Fig.~\ref{fig:Time}c, the calculated optical heating of Si substrate is shown for the case of fs-pulse train with inter-pulse distance 2~$\mu$s, revealing considerable and relatively slow heat accumulation after each pulse. Such an increase of background temperature is governed mainly by average intensity of incident light both inter-pulse duration and  

Also, it is important to mention that strong thermal localization in z-direction (i.e. in the case of $a \gg d$) yields much more efficient optical heating. Further optimization can be done with light source localization in lateral direction by reducing \textit{a}. However, the diffraction limit does not allow for strong and local optical heating, making it crucial to employ nanoparticles.

On this way, one of the most crucial differences between the optical heating of plain surface and nanoparticles is in importance of the optical losses. Indeed, on one hand, for the case of surface, higher absorption is almost always leads to stronger heating, whereas in resonant nanoparticles. In opposite, the most optimal cases for optical heating of the resonant particles can be achieved with low-loss materials, which we discuss in the next Sections.

\section{Optical heating of dielectric nanoparticles}

In order to understand the mechanism of resonant optical heating of optically resonant nanoparticle, one should study the light absorption cross-section of the nanoscale object. For this purpose, we consider light scattering by a spherical nanoparticle in homogeneous loss-free environment as basic primer widely utilized in  nanophotonics. The exact solution of this electromagnetic problem  was described by Gustav Mie back in the early XX$^{th}$ century, when he solved the problem on the plane wave scattering by a subwavelength sphere or infinite cylinder~\cite{mie1908beitrage}. The detailed solution of the Mie problem is provided elsewhere~\cite{bohren2008absorption}.

\subsection{Absorption of light by resonant nanoparticles}

\subsubsection{A brief summary of the Mie theory}

In general case, a dielectric spherical resonator  supports, so-called, Mie-modes of both electric and magnetic nature. The total scattering $C_{sca}$, extinction $C_{ext}$ and absorption $C_{abs}$ cross-section with predefined incident intensity $I_i$, wave vector $k$ are defined as follows 


\begin{align}\label{eq:cross-sections}
C_{sca} = \frac{2\pi}{k^2} \sum^{\infty}_{l=1} (2l+1) (|a_l|^2+|b_l|^2) \nonumber \\
C_{ext} = \frac{2\pi}{k^2} \sum^{\infty}_{l=1} (2l+1) Re(a_l+b_l) \nonumber \\
C_{abs} = C_{ext} - C_{sca},
\end{align}

where $a_{l}$ are the electric modes and $b_{l}$ are the magnetic ones, that are defined as:

\begin{align}\label{eq:mie-coefficient}
a_l = \frac{m\psi_l(mx)\psi'_l(x) - \psi_l(x)\psi'_l(mx)}{m\psi_l(mx)\xi'_l(x) - \xi_l(x)\psi'_l(mx)} \nonumber \\
b_l = \frac{\psi_l(mx)\psi'_l(x) - m\psi_l(x)\psi'_l(mx)}{\psi_l(mx)\xi'_l(x) - m\xi_l(x)\psi'_l(mx)},
\end{align}
and Riccati-Bessel functions are:
\begin{equation}\label{eq:43}
\psi_l(\rho) = \rho j_l (\rho) \qquad \qquad \qquad \xi_l(\rho) = \rho h^{(1)}_l (\rho).
\end{equation}

For the following coefficients $x$ is the diffraction parameter, $\rho = kr$, and $m$ is the relative refractive index.

\begin{equation}\label{eq:38}
x = ka = \frac{2\pi \textrm{n} a}{\lambda} \qquad \qquad \qquad m = \frac{k_1}{k} = \frac{\textrm{n}_1}{\textrm{n}}
\end{equation}

\begin{equation}\label{eq:23}
j_l(\rho) = \sqrt{\frac{\pi}{2\rho}} J_{l+1/2}(\rho)
\qquad \qquad \qquad
y_l(\rho) = \sqrt{\frac{\pi}{2\rho}} Y_{l+1/2}(\rho)
\end{equation}
$J_{n}(\rho)$ and $Y_{n}(\rho)$ are Bessel functions of first and second order.

\begin{figure}[ht!]
\centering
  \includegraphics[width=0.99\textwidth]{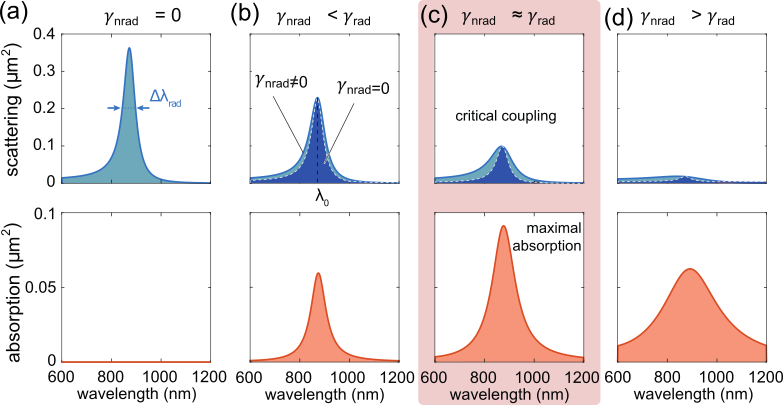}
  \caption{%
  \textbf{Radiative and Ohmic losses in dielectric Mie resonators.} Analytically calculated scattering (upper row) and absorption (lower row) cross-section through Mie theory for spherical nanoparticle of 215 nm diameter with fixed real part of the permittivity $\varepsilon{'} = 15.3$, which correspond to crystalline silicon in the visible range and imaginary part of the permittivity: (a) $\varepsilon{{''}} = 0$, (b) $\varepsilon{''} = 0.28$, (c) $\varepsilon{''} = 1$, and (d) $\varepsilon{''} = 4$.}
  \label{fig:mie_gamma}
\end{figure}

The typical scattering and absorption cross-section spectra  obtained with help of   Eq.~\ref{eq:cross-sections} are shown in Fig.~\ref{fig:mie_gamma}. The obtained spectra take into account only a single magnetic dipole (MD) eigenmode of a spherical nanoparticle of  215 nm diameter  corresponding to the coefficient $b_1$ only in Eq.~\ref{eq:mie-coefficient}. One can see that the spectra have a resonance at a wavelength $\lambda_0=2\pi c/  \omega_0$ corresponding to the fundamental MD Mie mode. The position of the resonance can be roughly estimated from the condition $2\pi R/\lambda_0\approx 1$.     While the real part of the permittivity $\varepsilon{'}$ is fixed at  15.3 which roughly corresponds to parameters of crystalline silicon in the visible range,  the imaginary part is varied: $\varepsilon{''} = 0$ (Fig.~\ref{fig:mie_gamma}a), $\varepsilon{''} = 0.28$ (Fig.~\ref{fig:mie_gamma}b), $\varepsilon{''} = 1$ (Fig.~\ref{fig:mie_gamma}c), and  $\varepsilon{''} = 4$(Fig.~\ref{fig:mie_gamma}d).  The increase of imaginary part $\varepsilon{''}$ provides the increase of non-radiative losses of the MD mode characterized by the rate $\gamma_{\rm nrad}$, while the radiative losses  channel is always present and its rate  $\gamma_{\rm rad}$ scales with the size  $\gamma_{\rm rad}/\omega_0\sim n^3$, where $n$ is the refractive index of the material.      

 The gradual increase of Ohmic losses $\gamma_{\rm nrad}$ results in broadening of the scattering spectra which is schematically shown in the upper row of Fig.~\ref{fig:mie_gamma} with red (radiative loss channel) and green (Ohmic losses channel) arrows. However, one can notice that the absorption spectra, which has the key importance form optothermal effects in nanostructures, has a non-monotonous dependence on the losses, while the scattering intensity gradually decreases.   At certain value of $\varepsilon{''}$, further increase of the nonradiative losses results in even decrease of the absorption cross-section. Therefore, the optimized regime is realized in Fig.~\ref{fig:mie_gamma}(c), which corresponds to maximal absorption cross section and so-called 'critical-coupling' regime provided by $\gamma_{\rm {rad}}$=$\gamma_{\rm rad}$ and can condition $C_{sca}=C_{abs}$ at the resonance as shown in Fig.~\ref{fig:mie_gamma}c. The conclusions in this sections correlated with those described in frameworks of other approaches~\cite{tribelsky2011anomalous, miroshnichenko2018ultimate}.  Despite that we have considered an example of a spherical particle, the coupling condition can be applied to any resonant nanostructure and its  generalized form will be considered below in Section \ref{sec:CMT}.
 
\subsubsection{Mechanism of heat generation} 

The energy absorbed by the nanostructures  goes for increasing its the temperature. The   density of the heat power $q(\mathbf{r})$ inside the NP can be derived basing on the Ohmic absorption light power:  

\begin{equation}
 q(\mathbf{r})=\frac{1}{2} \operatorname{Re}\left[\mathbf{J}^{\star}(\mathbf{r}) \cdot \mathbf{E}(\mathbf{r})\right]   
\end{equation}

where $\mathbf{J}(\mathbf{r})$ is the complex amplitude of the electronic current density inside the NP. As $\mathbf{J}(\mathbf{r})=i \omega \mathbf{P}$ and $\mathbf{P}=\varepsilon_{0} (\varepsilon(\omega)-1) \mathbf{E}$
one ends up with

\begin{equation}\label{eq:heatso}
   q(\mathbf{r})=\frac{\omega}{2} \operatorname{Im}(\varepsilon(\omega)) \varepsilon_{0}|\mathbf{E}(\mathbf{r})|^{2}=\dfrac 12 \sigma |\vec E(\vec r)|^2, 
\end{equation}

where the latter expression is obtained in  terms of the electric conductivity $\sigma$ owing to the relation between imaginary part of the permittivity determines the conductivity as $\sigma=\varepsilon_0\omega \text{Im}(\varepsilon)$, where $\varepsilon_0$ is the dielectric permittivity of vacuum. The total dissipated power $Q$ can then be expressed: 

\begin{align}\label{eq:Joule}
Q=\int_{V} q(\mathbf{r}) \mathrm{d}^{3} r
\end{align}

where the integral runs over the NP volume $V$. 

Generally, the larger the volume of an arbitrary nanoparticle, the stronger the heating.
In general case, integration in Eq.\ref{eq:Joule} over an arbitrary nanoparticle volume supporting eigenmodes allows to rewrite it in terms of an effective mode volume $V_{\rm eff}$ and spatially averaged field enhancement factor $<|E|>$, which determines how much energy can be accumulated inside the nanoparticle. Thus, we can rewrite Eq.\ref{eq:Joule} for the absorbed power as~\cite{zograf2017heat}:
\begin{equation}\label{temp_1}
Q \sim \sigma <|E|>^2\ V_{\rm eff}.
\end{equation}
From this expression, on can see that the effective mode volume inside the nanoparticle $V_{\rm eff}$ is very important for the optical heating of NP. Also, the effective mode volume can be very different from the physical volume of NP. For plasmonic nanoparticles (Re($\epsilon$)$<$0), the effective volume is defined by a skin depth, which is less than $\delta\approx20$~nm for most of metals in the visible range~\cite{palik1998handbook}. Thus, the effective mode volume of plasmonic nanoparticles is $V_{\rm eff}\approx\pi D^2\delta$. In opposite to metals, dielectrics support optical penetration depth much larger than the diameter of the nanoparticle in the visible range. The effective volumes of Mie-type modes are typically of the order of nanoparticle volume $V_{\rm eff}\approx \pi D^3 /6$. It means that the increase of the nanoparticle size is effective for the temperature increase in the case of dielectrics and less effective for the plasmonic nanoparticles.

On the other hand, in the case of a spherical geometry in homogeneous media, the absorbed power $Q$ can be calculated basing on the absorption cross-section (see Eq.\ref{eq:cross-sections}) from the Mie theory: 
\begin{equation}\label{eq:Qmie}
Q=C_{\mathrm{abs}} I_0
\end{equation}
This expression is quite useful for simple estimation of NP heating basing on the results of Fig.\ref{fig:mie_gamma}, as well as for temperature distribution analysis which we discuss below.

\subsection{Temperature distribution in nanostructures}

\subsubsection{Spherical nanoparticles in homogeneous media}

We now focus on the temperature distribution inside a nanostructure in the steady state regime under CW laser illumination. Assuming that $R$ is the typical  size of the nanostructure (sphere as a primer) and  thermal conductivity $\kappa$,  we consider that is  embedded in homogeneous media or placed on substrate with thermal conductivity $\kappa_s$. According to Eq.\ref{Eq:HeatT} the time required to reach the stationary regime in a nanostructure is of the order of $\sim c_p \rho R^2/\kappa$ which is  order of $~ 1\mu$s for $200$ nm particle. 

The temperature distribution can be obtained by solving Eq.\ref{Eq:HeatT} in the whole region of interest, which is valid as soon as diffusion mechanism of heat transport is applicable. Its violation occurs at the nanoscale when the mean free path of phonons $l$ becomes larger than the characteristic nanostructure size $l\gg R$. For the most materials of dielectric nanophotonics the phonon mean free path varies in the range 10-200 nm depending on the particular phonon models~\cite{chen2000phonon}, which is comparable with the typical size of  nanoparticles. Beyond the heat diffusion regime, one should consider Boltzman transport equation and possible ballistic phonon propagation, which is well summarized in the recent paper~\cite{cunha2020controlling}. The steady-state temperature distribution $T(\mathbf{r})$ inside and outside the NP can be, thus,  obtained as a solution  of the heat diffusion equation:

 
\begin{equation}
    \begin{array}{ll}{\nabla \cdot[{\kappa} \nabla T(\mathbf{r})]=-q(\mathbf{r})} & {\text { inside the } \mathrm{NP}} \\ {\nabla \cdot[{\kappa_S}\nabla T(\mathbf{r})]=0} & {\text { outside the } \mathrm{NP}}\end{array}
\end{equation}

The power density distribution $q(\ve r)$ can be strongly inhomogeneous as the field distribution is defined by the particular structure of the resonant mode. At the same time, the temperature appears to be almost constant across the nanoparticle volume~\cite{baffou2009heat} in case when the thermal conductivity of the nanoparticle material is much larger than the thermal conductivity of the environment $\kappa\gg\kappa_s$ (air, glass, liquid, etc.). Indeed, for a spherical NP of radius $R$, the calculations for homogeneous power density $q(\ve r)=Q/V$ leads to an analytical temperature dependence:

\begin{equation}
\begin{array}{ll}
T_{out} = \frac{Q}{4\pi\kappa_2r} + T_0, & {r>R} \\ T_{NP} = \frac{Q}{8\pi\kappa R}(1 - \frac{r^2}{R^2}) + \frac{Q}{4\pi\kappa_sR} + T_0, & {r<R}
\end{array}\label{eq:temp_solution}
\end{equation}
As one can see, the first term corresponding to temperature non-homogeneity vanishes as soon as  $\kappa_s \ll \kappa$. Finally, the nanoparticle temperature increase  $\Delta T_{\mathrm{NP}}$ can be simplified to the following expression:

\begin{equation}\label{eq:TempDep}
  \Delta T_{NP}=T_{NP}-T_0 \approx \frac{Q}{4\pi\kappa_sR}  = \frac{C_{\mathrm{abs}} I_0}{4\pi\kappa_sR}.
\end{equation}

As one can see from  Eq.~\ref{eq:TempDep}, that the temperature elevation is directly defined by the absorption cross section  $C_{\mathrm{abs}}$ reaching its maximum in critical coupling regime as shown in Fig.~\ref{fig:heat_mie}. 

Also, one can notice that  Eq.~\ref{eq:TempDep} resembles the expression for maximum temperature of surface heating Eq.~\ref{Eq:Tmax}, where nanoparticle's radius $R$ can be much smaller than the radius of the diffraction-limited focused Gaussian beam $a$. Moreover, the analytical consideration with Eq.\ref{eq:temp_solution} was done for the case of a spherical particle  in homogeneous surrounded by a uniform media. Nevertheless, it can be easily extended to other geometries by introducing  an effective conductivity parameter $\kappa_s$ in Eq.~\ref{eq:TempDep}, which depends on particular geometry of the structure. Baffou et al. have numerically extracted this effective parameter for a number of simple geometries of nanostructures geometries~\cite{baffou2010nanoscale}. 

\subsubsection{Effect of substrate: non-uniform thermal field distribution}

The thermal conductivity of solid materials utilized  in nanoplasmonics is normally much higher than of the environment one for  aqueous or gas surrounding. In this regard,  homogeneous thermal distribution within NPs becomes a good approximation in the most cases. However, this approximation for all-dielectric nanostructures is more critical and it may fail for  many particular geometries.  For example, if one considers crystalline silicon nanosphere optical heating in homogeneous aqueous media, under the assumptions that the thermal conductivity of silicon (156 W/(m$\cdot$K)) is much greater than the water's one (order of $\sim$1 W/(m$\cdot$K)), allows one to use the Eq.~\ref{eq:TempDep}. However, thermal conductivity of amorphous silicon is two orders of magnitude lower than of crystalline silicon 1.8 W/(m$\cdot$K) and is comparable to glass ($\sim0.8$ W/(m$\cdot$K)) often used as a substrate materials in nanofabrication process, while sapphire has even  much higher thermal conductivity of 34 W/(m$\cdot$K). Thus, the problem of thermal isolation between nanostructure and the substrate becomes  important for effective heating of all-dielectric nanostructures. 

\begin{figure}
\centering
  \includegraphics[width=.99\columnwidth]{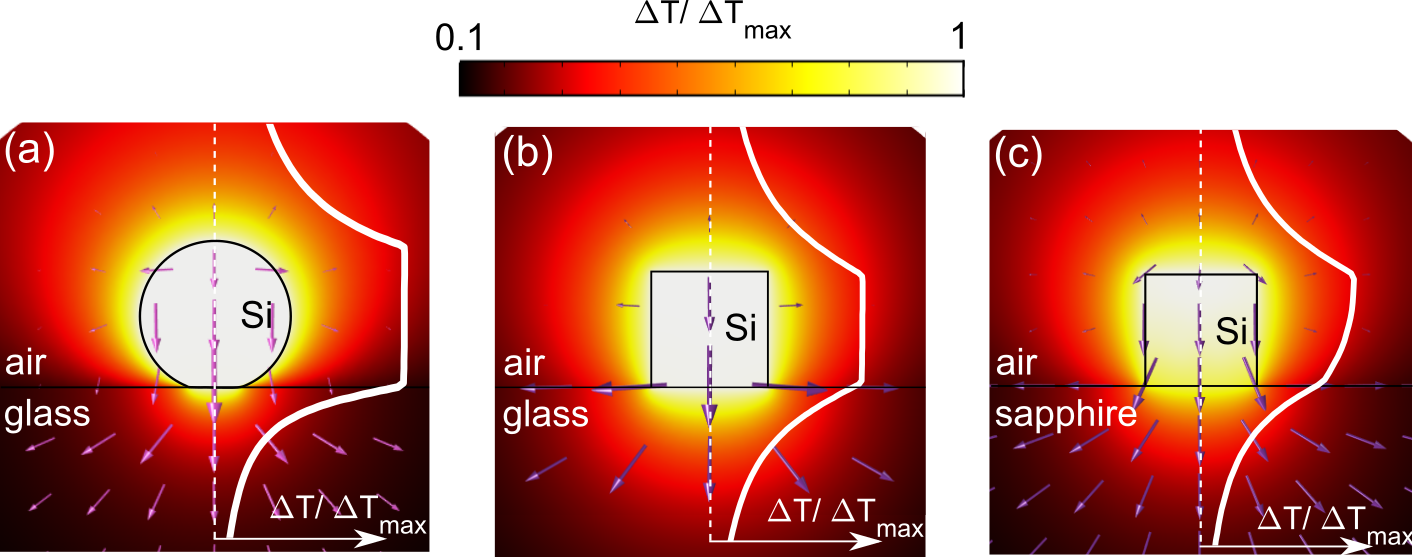}
  \caption{\textbf{Effect of substrate on temperature distribution in nanoresonators.} 
 Numerical simulation of the temperature distribution under CW laser heating of a single silicon (a) nanosphere on a glass substrate, (b) nanocylinder on a glass substrate and (c) nanocylinder on a sapphire substrate. The color-bar bar represents the scale with temperature increase normalized over maximum value of temperature distribution along the structure. White line corresponds to temperature profile along the linear section through the z-axis. Purple arrows demonstrate the heat flux. The arrow size is proportional to the flux magnitude.}
  \label{fig:distribution_substrate}
\end{figure}

Figure~\ref{fig:distribution_substrate} shows schematically the temperature distribution  inside nanoparticles placed over dielectric substrates. As one can see, the substrate material and the shape of the nanoparticle plays a significant role for thermal transfer under CW laser heating, as was already studied elsewhere~\cite{danesi2018using,danesi2018photo}. For two cases when  nanoparticle made of crystalline silicon of spherical and cylindrical shapes (Fig.~\ref{fig:distribution_substrate}a,b) the area of  thermal contact does not affect the homogeneity of temperature distribution inside the nanoparticle for glass substrate with low thermal conductivity. The opposite case occurs when the  substrate is made of sapphire with a significantly larger thermal conductivity. One can see in Fig.~\ref{fig:distribution_substrate}c, that the upper and bottom parts of the nanocylinder have distinguishable homogeneity in temperature values. For this particular reason, the color-bar is in logarithmic scale to demonstrate more clearly the presence of non-uniform temperature distribution pattern in such case.


\subsection{Effect of radiative losses on optical heating}
\label{sec:CMT}

Here, we provide the basics of nanoresonators heating. We start with a primer example of a resonators with a single resonant mode in the spectral region of interest. The mode of the resonator can be described within the coupled mode theory.  The field inside the resonator  $\ve E(\vec r, t)=a(t) \ve u(\vec r)/\sqrt{\veps_0}$, where $a(t)$ is the mode amplitude and $u(\vec r)$ is the mode spatial distribution. 
\begin{align}
\dfrac{d a(t)}{dt}=(-i \omega_0-\gamma)a(t)+i\sqrt{\Gamma_{rm}}f(t)
\label{eq:CMT}
\end{align}  
Here, $\omega_0$ is the resonant frequency of the mode, $\gamma=\gamma_{r}+\gamma_{nr}$ is the total losses of the mode consisting of radiative  ($\gamma_r$) and nonradiative ($\gamma_{nr}$) losses. The system is driven by the external field with mode amplitude $f(t)$ and $\Gamma_{rm}$ is the radiative losses of the resonator into this mode. Assuming a continuous wave excitation, we can consider the spectral representation of the problem $f(t)=\hat f(\omega)e^{-i\omega t}$ and  $a(t)=\hat a(\omega)e^{-i\omega t}$: 
\begin{align}
&-i\omega \hat a(\omega)=(-i \omega_0-\gamma)\hat a(\omega)+i\sqrt{\Gamma_{rm}}\hat f(\omega)\Rightarrow \nonumber \\
&\hat a(\omega)=\dfrac{i\sqrt{ \Gamma_{rm}}\hat f(\om)}{i(\om_0-\om)+\gamma}
\end{align}  

The optical heating of the nanostructure is provided by the ohmic losses of material. Indeed, the energy dissipated due to Joule  heating in the nanostructure can be estimated as follows: 
$$
Q=\int_{V}\sigma |E(r)|^2 dV=|a(t)|^2\int_V \sigma |u(r)|^2 dV,  
$$
where $\sigma$ is the conductivity of material. The total energy losses can be derived from \eqref{eq:CMT} relying on $Q_{tot}=d|a|^2/dt=-(\gamma_r+\gamma_{nr})|a|^2$, thus $Q=\gamma_{nr}|a|^2$, which immediately provides us with the expression for nonradiative losses $\gamma_{nr}=\dfrac{1}{\veps_0}\int_V \sigma |u(r)|^2 dV$. 


\begin{figure}
\centering
  \includegraphics[width=0.95\textwidth]{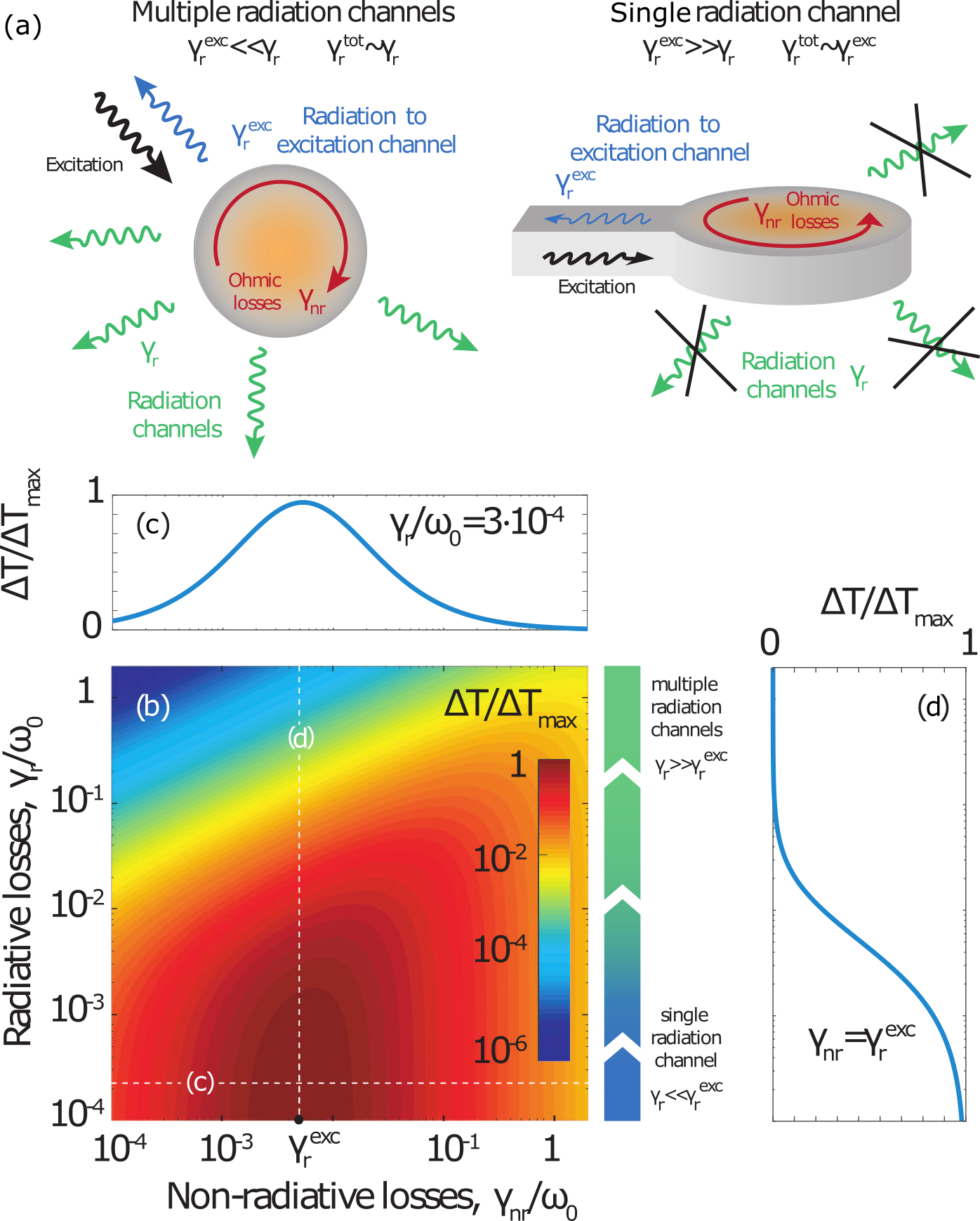}
  \caption{%
  (a) Different excitation channels and (b) calculated losses for the optical heating optimization. The excitation channel coupling constant is taken $\Gamma_{rm}/\omega_0=5\cdot 10^{-3}$. One-dimensional plots (c) and (d) correspond to cross-sections of the map along the white dash lines.}
  \label{fig:coupling}
\end{figure}

The temperature, which acquires  the resonator in the stationary regime  is homogeneous across its volume and has a simple expression: 

\begin{align}
\Delta T=\dfrac{Q}{4\pi \kappa_s R}=\dfrac{1}{4\pi \kappa_s R}\dfrac{\gamma_{nr}{\Gamma_{rm}}|\hat f(\om)|^2}{(\om_0-\om)^2+(\gamma_r+\gamma_{nr})^2} ,
\end{align}
where $\kappa_s$ is the thermal conductivity of the surrounding media, and $R$ is the nanoparticle radius.  The maximal  temperature can be achieved at the resonance: 
$$
\Delta T_{max}=\dfrac{1}{4\pi \kappa_s R}\dfrac{\gamma_{nr}{\Gamma_{rm}}|\hat f(\om)|^2}{(\gamma_r+\gamma_{nr})^2} 
$$
For a case of a plane wave the amplitude $\hat f$ is related to the field  intensity as $\hat f=\sqrt{c_0\veps_0 \omega  R^2} E_0 $.

{\bf Critical coupling.} One can see that the maximal temperature is limited by the losses factor but also defined by the ratio between radiative and non-radiative losses, which is often referred to as a critical coupling regime: the maximal heating of the structure will be observed one radiative and non-radiative losses will be balanced $\gamma_r= \gamma_{nr}=\gamma_{opt}$. In the critical coupling regime, the maximal temperature will inverse proportional to the overall losses $\Delta T_{max}\sim1/\gamma_{opt}$, clearly showing that with the decrease of the total losses the heating efficiency will be monotonically increased.

{\bf Single radiative channel.} Another important issues is that  the heating efficiency is also proportional to the far-field coupling constant $\Gamma_{rm}$. Open subwavelength nanophotonic structures are usually coupled to many radiative channels, providing that $\gamma_r=\gamma_r'+\Gamma_{rm}$ and $\gamma_r'\gg \Gamma_{rm}$ (see Fig.~\ref{fig:coupling}). However, in particular cases the number of radiative channels can be very limited and even only one, thus making $\gamma_r'=0$. In this limiting case  (see Fig.~\ref{fig:coupling}) the total radiative losses are associated with the excitation channel only $\gamma_r= \Gamma_{rm}$ and then the maximal temperature in the critical coupling regime will be equal to $\Delta T_{max}=|\hat f(\om)|^2/{(16\pi \kappa_s R)}$.

\subsection{Effect of nonradiative losses on optical heating}

The origin of optical losses in non-plasmonic materials in visible and infrared region is mainly related to the mechanism of inter-band or intraband electron absorption. While the former are related to valence-to-conductance band transitions or to excitonic transitions, the latter are connected to absorption of light by free carriers present in semiconductors.

\subsubsection{Losses in natural materials}

\begin{figure}
\centering
  \includegraphics[width=.99\textwidth]{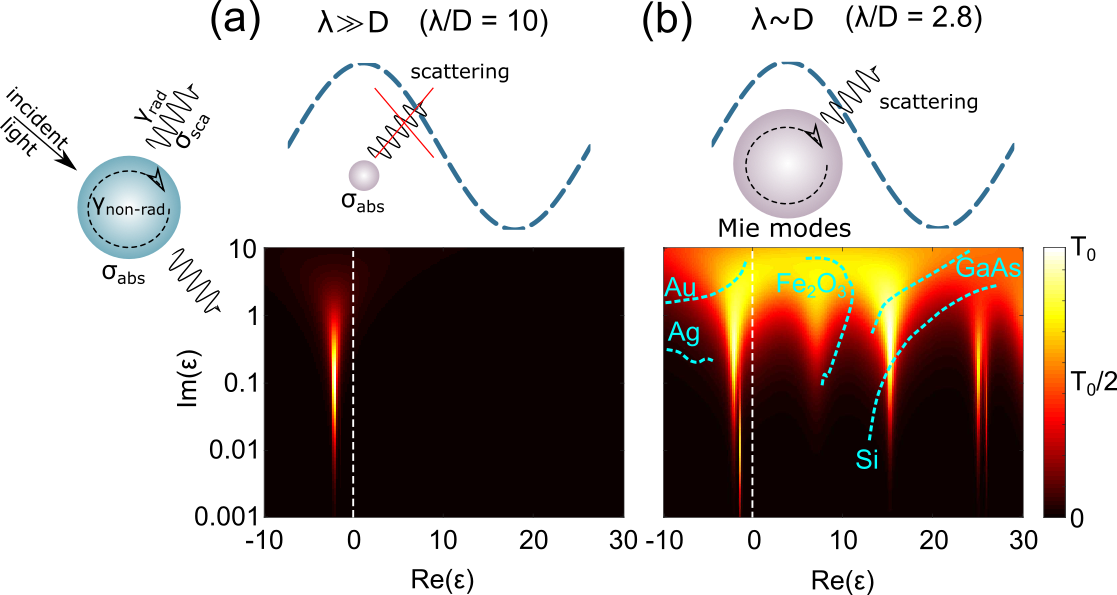}
  \caption{\textbf{Effect of complex dielectric permittivity on heating.}
   Theoretically calculated (from Mie theory) heating maps for spherical nanoparticles with deeply subwavelength (a) ($\lambda$)/(D)=10 and near-wavelength (b) ($\lambda$)/(D)=2.8 diameters for different real and imaginary parts of permittivity in homogeneous medium (air). Insets on top of the heating maps schematically show the size of the nanoparticle comparing to the excitation wavelength and depicts the mechanism of light-structure interaction. The scale of an imaginary part of permittivity is logarithmic. The white dashed vertical lines correspond to zero permittivity (metal-dielectric transition) and cyan dashed lines correspond to Im($\varepsilon$) and Re($\varepsilon$) values for given materials in the visible range. (adopted from~\cite{zograf2017resonant}) Left inset: Schematic representation for radiative and non-radiative losses in a nanoresonator at electromagnetic wave excitation.}
  \label{fig:heat_mie}
\end{figure}

The results of the calculations basing on Eq.~\ref{eq:TempDep} are shown in Fig.\ref{fig:heat_mie}, where the heating of dielectric and metallic nanoparticles of different sizes are compared. Namely, the dependence of temperature increase  inside a nanoparticle with defined real (Re($\epsilon$)) and imaginary (Im($\epsilon$)) parts of permittivity is presented. These results give general conclusion that the relatively large dielectric nanoparticles can be heated as effective as plasmonic ones, whereas their Im($\epsilon$) can be significantly smaller. Indeed, Figure~\ref{fig:heat_mie} shows that low optical losses inherent for most of dielectrics do not necessarily result in weak photo-induced heating of nanoparticles. 

For very small nanoparticles where $\lambda \gg D$ one can see significant optical heating in the region of negative $Re(\varepsilon)$, which means only metals support heating in small structures. However, upon increasing the diameter of the nanoparticle, one can observe efficient optical heating of dielectric with even relatively low amount of optical losses $Im(\varepsilon)$. If the size of the structure reaches the order of the incident wavelength and larger, optical heating mechanism becomes similar to the absorption governed by the Beer's law for bulk samples, except the cases of excitation of high-order optical modes in perfectly shaped resonators.

From the Eq.~\eqref{temp_1} it is clear that the increasing Ohmic losses does not necessary lead to the rise of the light absorption by the nanoparticle. Basing on the performed calculations, we stress that the dielectric nanoparticles can be efficiently heated by light illumination when their radiative losses are equal to the Ohmic ones. Since dielectrics have low nonradiative losses, this condition is fulfilled for relatively big nanoparticles. Plasmonic nanoparticles, on the contrary, are expected to show the most effective heating for significantly smaller sizes~\cite{govorov2007generating}.

\subsubsection{Doping of resonant dielectric nanoparticles}

\begin{figure}[t!]
\centering
  \includegraphics[width=.99\textwidth]{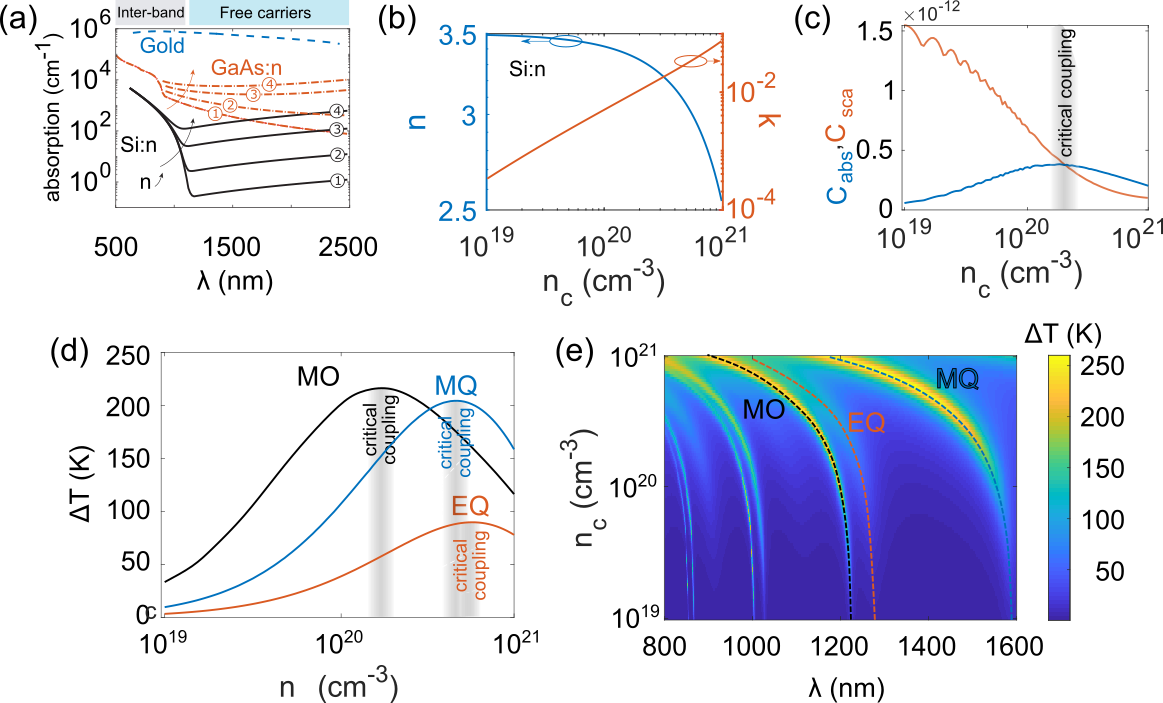}
  \caption{\textbf{Loss engineering via doping.} (a) Light absorption in Au, c-Si:n, and GaAs:n. The absorption curves for different level of doping are shown: 1 -- $10^{17}$ cm$^{-3}$; 2 -- $10^{18}$ cm$^{-3}$; 3 -- $10^{19}$ cm$^{-3}$; 4 -- $5\cdot 10^{18}$ cm$^{-3}$. (b) Real (blue) and imaginary (red) parts of refractive index as a function of doped free carriers concentration. (c) Maximum values of scattering (red) and absorption (blue) cross-sections in 800 to 1600 nm excitation wavelength at different doping levels for magnetic octupole (MO) mode only. (d) Optical heating temperature of single spherical nanoparticle with 315 nm radius with refractive index defined by b) and c) as a function of doping level. (e) Optical heating of spherical nanoparticle as a function of doped carriers concentration at a certain optical mode - MO, MQ, EQ - magnetic octopole, magnetic quadrupole, electric quadrupole modes, respectively. 
}
  \label{fig:doped_sphere}
\end{figure}

A beneficial property of non-plasmonic materials is their ability for tuning the optical losses. Though the inter-band losses are normally provided by the band structure of the solid and hardly can be varied, the excitonic absorption can be tuned in the wide range by tuning the excitonic transition is perovskite materials~\cite{tiguntseva2018tunable,tiguntseva2018light,akkerman2015tuning}. More conventional methods of intraband absorption tuning  require doping - intentional introduction of impurities into an intrinsic semiconductor for the purpose of modulating its electrical, optical and structural properties. Thus allowing  effective alternation of the band structure~\cite{erwin2005doping,spear1975substitutional}. In Fig.\ref{fig:doped_sphere}a the spectral dependencies of the optical  absorption in c-Si and GaAs are shown for different values of the doping level. One can see that for the short wavelength region the interband indirect optical transition govern the optical losses and are almost independent on the doping level. For the photon energy below the band gap the intraband free carrier absorption becomes dominant and doping drastically increases the absorption. Despite of that in the whole spectral region of the interest the optical losses of gold prevail over the losses in the semiconductors. The  doping of semiconductors also results in the change of the real part of the refractive index, which stays relatively low in the visible near-IR spectral regions. The ability to tune the losses can be efficiently used for controlling the optical heating of the nanostructure, which will be discussed in detail below.  

One of the possible ways to vary nonradiative losses of a nanoresonator, thus the nonradiative losses of an optical mode, is to dope the nanoresonator material before fabrication with free carriers. Doping of a single conventional semiconductor nanoparticles is still quite a challenging task except some nanostructures based on hybrid halide perovskites, where one can achieve a drastic change in the optical and conductive properties by in-situ nanoparticles doping~\cite{tiguntseva2018tunable}. In this regard, for the doping one should consider the films which subsequently would be reshaped by lithography techniques into metasurfaces and nanoparticles. Thus, a resonant silicon nanodisk is a reasonable structure for consideration.

To start with, it is necessary to estimate the affection of doping on optical and conductive properties of the material, since it impacts significantly on optical properties~\cite{lewi2015widely}. The free carrier contribution to the semiconductors permittivity is described by a Drude model ~\cite{kittel1976introduction}:
\begin{gather}\label{epsilon}
\varepsilon ' = \varepsilon_{\infty}\Big(1~-~\frac{\omega_{p}^2 \tau^2}{1+\omega^2 \tau^2}\Big) \\
\varepsilon'' = \frac{\varepsilon_{\infty}\omega_{p}^2 \tau}{\omega(1+\omega^2 \tau^2)}~,
\end{gather}
where the plasma frequency $\omega_{p}$ and scattering time $\tau$ are defined as $\omega_{p} = \sqrt{\frac{Ne^2}{m_c\varepsilon_{\infty}\varepsilon_{0}}}$ and $\tau = \frac{\mu m_c}{e}$, where $N$ is the free carrier concentration, $e$ is the electron
charge, $m_c$ is the conductivity effective mass and for n-type doping of silicon $m_c=0.26\cdot m_e, m_e$ is the electron mass, $\varepsilon_{0}$ and $\varepsilon_{\infty}$ are the permittivity of free space and
the high frequency permittivity, respectively, and $\mu$ is the free carrier mobility. 


The electron mobility and hole mobility have a similar doping dependence: for low doping concentrations the mobility is almost constant and is primarily limited by phonon scattering. At higher doping concentrations, the mobility decreases due to ionized impurity scattering with the ionized doping atoms. The actual mobility also depends on the type of dopant.
    
 The mobility at a particular doping density is obtained from the following empiric expression:
\begin{equation}
\mu = \mu_{min} + \frac{\mu_{max} - \mu_{min}}{1+\big(\frac{N}{N_r}\big)^{\alpha}}~,
\end{equation}
where fitting parameters for phosphorous doping of silicon $\mu_{min} = 68.5~cm^2/V\cdot s$, $\mu_{max} = 1414~cm^2/V\cdot s$, $N_r = 9.2\cdot10^{16}~cm^{-3}$, $\alpha = 0.711$.

The mobility of the carriers and concentration affects the permittivity of the material according to the Eq.~\ref{epsilon}. Slight change of the real part of permittivity affects the spectral position and quality factor of the resonance as one can see in Fig.~\ref{fig:doped_sphere}(d) where temperature of the nanosphere of 315nm radius under plane wave illumination (as schematically depicted in Fig.~\ref{fig:doped_sphere}(a) is shown. The imaginary part of refractive index defines the absorption in bulk material and, therefore, optical heating. One can notice blueshift of the excited optical modes with increase of the doping level. This occurs due to decrease of the refractive index as one can see from Fig.~\ref{fig:doped_sphere}(b). On the other hand, the increase of the free carriers with doping level increases dramatically the imaginary part of the refractive index, thus lowering the Q-factor of the mode, therefore spectrally broaden the resonance. However, after reaching certain level of carriers concentration, further doping does not boost optical heating, therefore it means that higher nonradiative losses does not necessarily lead to enhanced absorption and optical heating.

Explicitly the latter one is shown in Fig.~\ref{fig:doped_sphere}(e) where optical heating at certain optical modes MO, MQ, EQ (magnetic octopole, magnetic quadrupole, electric quadrupole modes, respectively) for different levels of doping is shown. Every particular optical modes is described by its radiative losses $\gamma_{rad}$ and non-radiative, which are mostly Ohmic losses due to Joule heating $\gamma_{Ohmic}$. The latter one is defined by imaginary part of refractive index, therefore by doped carriers concentration. In general, the higher the optical mode order, the better its quality factor and lower the radiative losses $\gamma_{rad}$. Thus, the radiative and non-radiative losses balance occur at lower doping concentrations. Indeed, for higher order and higher Q-factor MO mode, the optimal optical heating is being reached at lower doping levels, whereas for MQ and EQ losses match at higher doping concentration. This result is consistent with previous predictions for plasmonic~\cite{tribelsky2011anomalous} and all-dielectric~\cite{miroshnichenko2018ultimate} nanoparticles described by Prof. Tribelsky and co-authors. The so-called ultimate absorption regime is being realized, where absorption matches scattering as manifested in~\cite{miroshnichenko2018ultimate}, therefore the most efficient optical heating occurs at the same conditions. The Fig.~\ref{fig:doped_sphere}(c) depicts the maximum scattering cross-section and the maximum absorption cross-section of a single spherical nanoparticle of 315 nm radius at different values of doping for MO contribution only.


\subsubsection{Loss control via multiphoton absorption}

In nonlinear optics, it is well known effect when optical absorption becomes a function of incident light intensity~\cite{boyd2020nonlinear}. In this case, several photons initiate an interband transition and generation of free carriers in the conduction band of dielectric. Once this process involves the transitions through virtual states in the band gap, it becomes considerable at relatively high light intensities. 
The most often case in semiconductors pumped by near-IR light is two-photon absorption (TPA), which was observed in various designs and materials at intensities $>$1~MW/cm$^2$. Generally, total absorption ($\alpha$) with linear $\alpha_0$ and non-linear $\alpha_{NL}$ parts is written as 
\begin{equation}\label{TPA}
    \alpha = \alpha_0 + \alpha_{NL} = \alpha_0 + \beta I,
\end{equation}
where $\beta$ is the two-photon absorption coefficient, which is directly connected with third-order permittivity $\chi^{(3)}$ as $\chi^{(3)} = (c\beta\varepsilon)/(8\pi\omega)$, $\varepsilon$ is the linear dielectric permittivity, $\omega$ is the radial frequency of light, and $c$ is the speed of light in vacuum. However, in dielectric nanoparticles, TPA should be taken into account in Equation~\ref{Eq:carriers} via free carriers generation rate written as the following terms~\cite{baranov2016nonlinear}:
\begin{equation}\label{TRAGen}
G_e = \frac{Im(\varepsilon)}{8\pi\hbar}|E_{in}|^2  + \frac{Im(\chi^{(3)})}{16\pi\hbar}|E_{in}|^4 
\end{equation}
where $E_{in}$ is the averaged electric field over the nanoparticle volume calculated from the Mie coefficients~\ref{eq:cross-sections}-\ref{eq:mie-coefficient}. 

As a result, solving the Equations~\ref{Eq:TTM}-\ref{Eq:gamma} together with \ref{eq:cross-sections}-\ref{eq:mie-coefficient} one can reveal the critical coupling conditions for nonlinear light absorption. Remarkably, that in the infrared range, TPA is often much stronger than linear absorption, and, thus, unexpected reaching the most optimal conditions for high quality factor resonances can result in the nanostructures damage as compared with anticipated damage threshold for bulk materials. On the other hand, such prompt processes as TPA open the new avenues for ultrafast optical switching by heating of resonant dielectric nanostructures.

\subsection{Variation of physical properties at elevated temperatures}

As one can see from Fig.~\ref{fig:distribution_substrate}, substrate materials with different thermal conductivities can dramatically affect both the heating temperature and the temperature profile. It should be noted, that upon reaching elevated temperatures, the thermal properties of materials can drastically change. Such approach is well described by Y. Li $\textit{et al.}$~\cite{li2021transforming} through macro- and micro-engineering of the structures and materials.
\begin{figure}[h!]
\centering
  \includegraphics[width=.45\columnwidth]{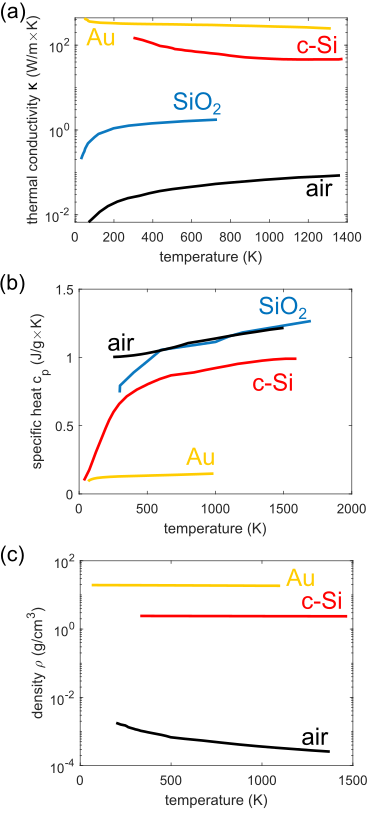}
  \caption{\textbf{Thermal physical properties at elevated temperatures.} 
(a) Thermal conductivity for Au~\cite{powell1966thermal}, c-Si~\cite{shanks1963thermal}, a-SiO$_2$~\cite{cahill1990thermal} and air~\cite{stephan1985thermal}. (b) Specific heat for Au~\cite{takahashi1986heat}, c-Si~\cite{okhotin1972thermophysical}, SiO$_2$~\cite{chase1998data} and air~\cite{hilsenrath1955tables} . (c) Density of c-Si, Au and air at elevated temperatures.}
  \label{fig:properties}
\end{figure}

\subsection{Nonuniform near-field distribution} 

Nanoantennas based on either plasmonic or all-dielectric materials can serve efficiently for electromagnetic field localization at the nanoscale, thereby can significantly enhance the field across small volumes~\cite{lim2010nanogap, bakker2015magnetic, caldarola2015non, nam2016plasmonic}. One should distinguish two main cases in near-field localization: \textit{external} and \textit{internal} relatively to the material of NPs. The first one corresponds to the localization of the field outside the nanostructure near its surface, whereas the second one accumulates energy inside the nanoobject. 

\begin{figure}[h!]
\centering
  \includegraphics[width=.95\columnwidth]{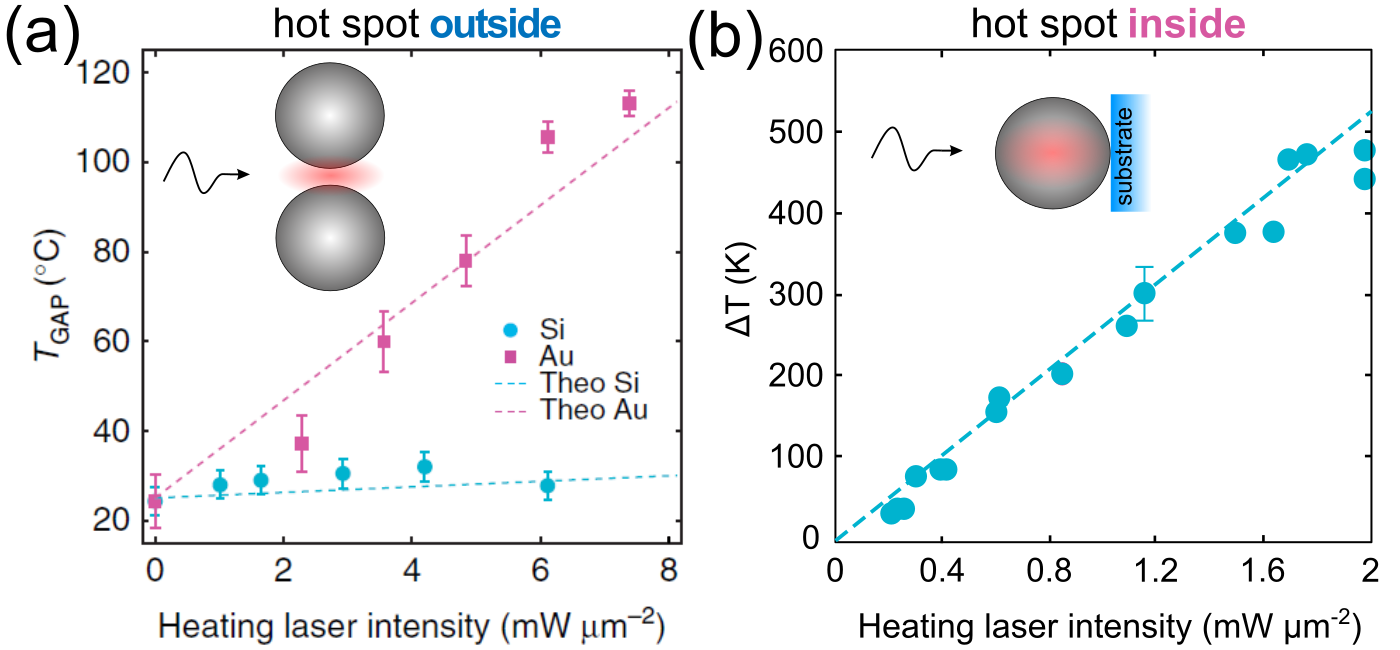}
  \caption{\textbf{Cold and hot regimes of light interaction with nanoparticles.} 
 (a) Extracted temperature in the gap for selected silicon (cyan) and gold (magenta) nanoantennas as a function of the heating laser intensity at 860~nm. The dashed lines show the numerical calculations for the temperature at the gap, presenting good agreement with the experimental data.~\cite{caldarola2015non}. (b) Experimental data (red circles) and numerical calculations (solid lines) for a spherical
silicon nanoparticle with diameter 350~nm on glass.~\cite{zograf2017resonant}}
  \label{fig:Hotspots}
\end{figure}

The first approach is beneficial for the cases when one aims to avoid overheating of the design and provide sensing experiments with the materials deposited on the nanostructure. In this case, employing all-dielectric nanophotonic designs has some advantage over metallic-based ones as shown in Fig.~\ref{fig:Hotspots}a adopted from~\cite{caldarola2015non}. All-dielectric nanostructures with external near-field enhancement can be also designed as nanotips~\cite{gervinskas2013surface, lagarkov2017light, mitsai2018chemically} or metasurfaces supporting optical modes with high quality factors~\cite{tittl2018imaging, yesilkoy2019ultrasensitive}.

Designing the hot-spots inside the material of nanoparticles and initiation of efficient optical heating is prospective for many applications discussed in Section~5.

\newpage
\section{Thermometry with all-dielectric nanoantennas}

 
Measuring temperature at the nanoscale is a challenging problem in modern nanoscale science.~\cite{brites2012thermometry,bradac2020optical} Generally, an all-optical approach based on temperature-sensitive optical response of nanostructures is likely less invasive than one requiring the reading of an electrical signal, making it more suitable for uses such as measuring temperature inside a living cell of living tissue. In this regard, the all-dielectric photonics gives additional tools based on the light emission from the non-metallic material itself rather than from any additional external markers commonly  utilized in thermoplasmonics.

Nanothermometry with nanoantennas can be done in different ways. As schematically shown in Fig.~\ref{fig:boxthermo}a, thermal sensitivity can arise from the material of resonant nanoparticle or from the surrounding material.

\subsection{Raman scattering} 

Generally, Raman response is inelastic scattering of incident light on crystal lattice phonon, thus higher the purity and cristallinity of the nanostructure - higher the quality factor and intensity of the Raman response. The Raman scattering allow to measure temperature either to the spectral shift of the Stokes signal~\cite{balkanski1983anharmonic, zograf2017resonant} or by Stokes/anti-Stokes ratio~\cite{hagiwara2018co2}. 

\begin{figure}
\centering
  \includegraphics[width=.79\columnwidth]{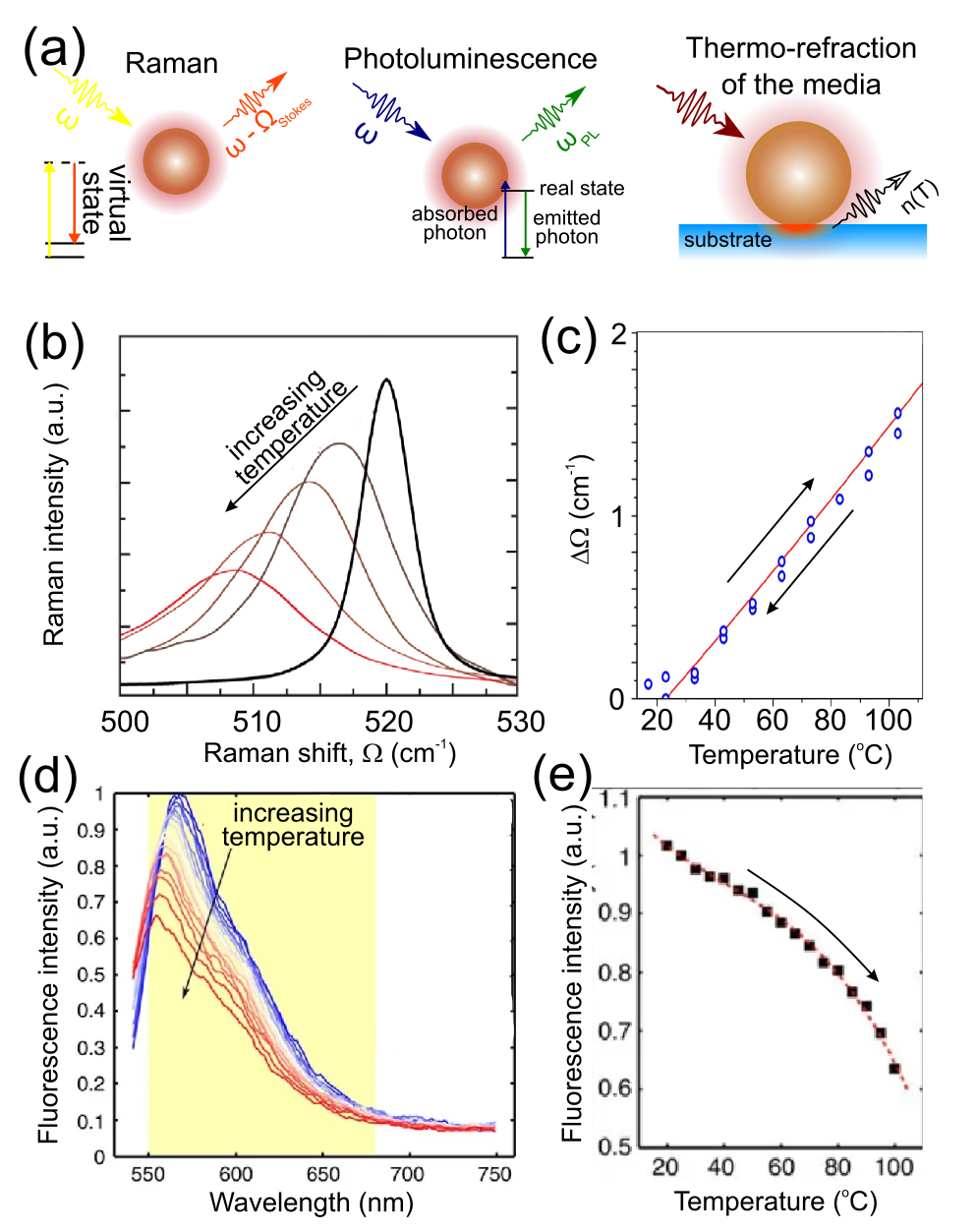}
  \caption{ \textbf{Nanothermometry with all-dielectric nanoantennas.}
 (a) Schematic illustration of nanothermometry with thermally sensitive nanoparticle only (left) and with additional thermal sensitivity from surrounding material. (b) Raman spectra of c-Si NPs at different intensities~\cite{zograf2017resonant}. (c) Connection between Raman shift and temperature as well as the fit used as a calibration curve to extract the corresponding temperature values. (d) Nile Red emission spectra taken at different temperatures. The marked zone shows the detection spectral range (550-680~nm). (e) The integrated intensity in the detection spectral range as well as the fit used as a calibration curve to extract the corresponding temperature values.
}
  \label{fig:boxthermo}
\end{figure}

Unlike metals, crystalline dielectrics and semiconductors possess pronounced Raman signal at room temperature, making it possible to provide direct Raman nanothermometry during optical heating even of single nanoparticle as shown in Fig.\ref{fig:boxthermo}b for the case of silicon. Indeed, the spectral position of a Raman line is known to be thermo-sensitive due to anharmonic and multiphonon-interaction effects in lattice vibrations.~\cite{balkanski1983anharmonic} Therefore, the frequency of an optical phonon ($\Omega$) responsible for Raman signal is dependent on temperature as
\begin{equation}\label{eq:Raman}
\begin{aligned}
  \Omega (T) = \Omega_0 + A\Bigg(1 + \frac{2}{e^x - 1}\Bigg) + \\ 
  + B\Bigg(1 + \frac{3}{e^y - 1} + \frac{3}{(e^y-1)^2}\Bigg),
\end{aligned}
\end{equation}

where $\Omega_0$ is photon frequency at zero temperature, \textit{A} and \textit{B} are constants, $x~=~{\hbar \Omega_0}/{2kT}, y~=~{\hbar \Omega_0}/{3kT}$, which works for crystalline silicon~\cite{balkanski1983anharmonic}. This function is plotted in Fig.~\ref{fig:boxthermo}c fitting experimental values calibrated with using standard heat plate. However, dielectric nanoparticles often are not monocrystalline and consist of nanoscale grains affecting (spectral shift and broadening) the Raman signal. Careful comparison of the Raman spectral shift fittings for different silicon samples based on the Eq.~\ref{eq:Raman} and simpler linear approximation were carried out in the work~\cite{karpinski2020optical}, revealing considerable differences between silicon bulk wafer and nanoparticles.

The temperature resolution of Raman-based method is fundamentally limited by the spectral width of the phonon line ($\sim$1~cm$^{-1}$), and some signal post-processing could be applied to extract information on temperature more precisely.

Another Raman nanothermometry approach is based on the comparison of Stokes and anti-Stokes signal intensities. It is known that anti-Stokes/Stokes Raman signals ratio increases with growth of the temperature of crystallized dielectrics (e.g. silicon~\cite{hart1970temperature}). Because of Boltzman distribution of phonons population, an exponential dependence on temperature is observed for this ratio:
\begin{equation}\label{eq:antiStokes}
  \frac{I_A}{I_S}=e^{-\frac{\hbar\omega_0}{kT}},
\end{equation}
where $I_A$ and $I_S$ are intensities of anti-Stokes and Stokes respectively, $\hbar$ is reduced Planck constant, $\omega_0$ is optical phonon frequency of silicon, $k$ is Boltzmann constant and $T$ is temperature. 

Remarkably, Raman signal can be strongly enhanced at optical resonances in all-dielectric designs~\cite{dmitriev2016resonant, alessandri2016enhanced, baranov2018anapole, zograf2020stimulated, raza2020raman}. However, this property can reduce an accuracy of such intensity-based method as Stokes/anti-Stokes thermometry at high temperatures, because of thermo-refractive effects changing resonant properties of dielectric nanostructures and, thus, efficiencies of Raman scattering at different wavelengths in a different manner. 

\subsection{Photoluminescence}

The materials like dielectrics possessing band-gap and supporting direct interband transitions usually possess efficient photoluminescence (PL). PL process starts when material of nanostructure absorbs incident light with subsequent electron-hole pair generation, relaxation to the bottom of conduction band (or exciton formation), and further re-emission of photon with energy of the band gap (or excitonic energy). PL contains information on the emitting material temperature, because
PL spectral position ($\hbar\omega_{PL}$), life-time ($\tau_{PL}$), and quantum yield ($QY_{PL}$) are dependent on temperature as
\begin{equation}
   \hbar\omega_{PL}(T) \sim E_g(T) = E_g(0)-\dfrac{\xi_1 T^2}{T+\xi_2},
\end{equation}
\begin{equation}
    \frac{1}{\tau_{PL}(T)}= k_{tot} = k_{rad}+k_{nrad}(T)
\end{equation}
\begin{equation}
    QY_{PL}(T) = \frac{k_{rad}}{k_{tot}} = \frac{k_{rad}}{k_{rad}+k_{nrad}(T)}
\end{equation}
where $E_g$ is the band gap, $k_{rad}$ and $k_{nrad}$ are radiative and nonradiative recombination rates, respectively. In turn, $k_{nrad}$ is temperature-sensitive following Arrenous-like behaviour $k_{nrad}\sim k_{nrad,0}e^{-\delta E/k_BT}$, where $\delta E$ is the constant the activation energy for the nonradiative process and $k_B$ is the Boltzmann constant. $\xi_1$ and $\xi_2$ are material dependent constants.

Additionally, linewidth of the light-emitting system is temperature-dependent parameter. For instance, in various excitonic materials homogeneous broadening is due to scattering of the excitons by optical phonons and acoustic phonons~\cite{rudin1990temperature}:
\begin{equation}
    \Gamma(T)=\sigma T+\Gamma_{LO}\left[e^{(E_{LO}/k_B T)}-1\right]^{-1},
\end{equation}
where $\sigma$ is the exciton–acoustic phonon coupling coefficient, $\Gamma_{LO}$ is the exciton–longitudinal optical (LO) phonon coupling coefficient, $E_{LO}$ is the LO-phonon energy.

As shown in Fig.~\ref{fig:boxthermo}d,e, PL of dye enhanced by silicon nanoantennas is getting quenched exponentially with increase of temperature owing to the growth of non-radiative recombination rate in the dye molecules~\cite{caldarola2015non}. The achieved resolution is around few degrees. Also, thermal shift of PL line can be observed in Fig.~\ref{fig:boxthermo}d, but it can not be extracted owing to filtering of the short-wavelength wing. Thermally-sensitive PL of WS$_2$ flakes coupled with Si nanoparticles were also employed for nanothemometry with resolution around 5--10~K.\cite{yan2020active}

\subsection{Nonlinear scattering}

Silicon nanoparticles placed on phase changing materials (e.g. VO$_2$) demonstrated considerably tunable scattering spectra owing to strong variation of optical properties of their surrounding media,~\cite{yan2020active} yielding temperature resolution around 1~K. Because of high thermo-refractive coefficients for such materials as silicon, light scattering from Si nanoparticles at high temperatures can be also used for nanothermometry with high spatial resolution.~\cite{duh2020giant} In contrast to plasmonic nanoparticles, which are also demonstrated thermal-dependent scattering behavior due to strong increase of imaginary part of metal dielectric permittivity~\cite{sivan2017nonlinear}, the all-dielectric approach might have higher potential because of more degrees of freedom related to thermal tunability of far-field response caused by the interplay between magnetic and electric Mie-like resonances.

\subsection{Comparison of different nanothermometers}

\textbf{Signal level and acquisition time.}
Efficiency of the light-emission process is one of crucial parameters because it determines the applicability of the nanothermometry based on this process. Indeed, low-efficient emission would require highly-sensitive detectors, long acquisition times, and expensive detection optical schemes.~\cite{carattino2018gold} Typically, photoluminescence is several orders of magnitude more efficient process as compared to Raman scattering, while elastic scattering is strongly dependent on optical contrast. However, the defect-intolerant materials can possess low PL quantum efficiency owing to high defect concentration and, thus, fast nonradiative channels of electrons relaxation. In this regard, light-emitting nanoantennas made of defect-tolerant materials can be good candidates for nanothermometry applications.~\cite{tiguntseva2018light} Moreover, achieving lasing regime in nanoantennas would accelerate radiative recombination via stimulated mechanism, making the quantum yield almost maximum.~\cite{tiguntseva2020room, mylnikov2020lasing} On the other hand, Raman scattering can be enhanced by several orders of magnitude at Mie resonances,~\cite{dmitriev2016resonant,baranov2018anapole} making the signal acquisition time as fast as several seconds. As a result, PL, Raman, and elastic scattering from resonant all-dielectric nanostructures can give comparable seconds-level signal strongly dependent on various conditions.

\begin{figure}[t!]
\centering
  \includegraphics[width=.59\textwidth]{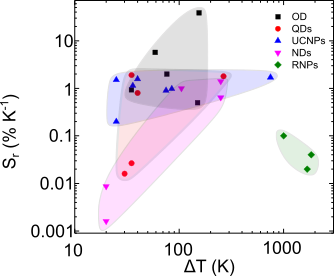}
  \caption{\textbf{Comparison of nanoscale thermometers.} Experimental parameters relative sensitivity $S_r$ and working temperatures range $\Delta$T ) for the known nanothermometers summarized in Table 1 (dots). Organic dyes (OD, black), quantum dots (QDs, red), upconversion nanoparticles (UCNPs), nanodiamonds (NDs), resonant nanoparticles (RNPs).
}
  \label{Thermo_chart}
\end{figure}

\textbf{Sensitivity.}
Sensitivity is an absolute quantity: it specifies the smallest, absolute amount of change that can be detected by the sensor.  in general, different nanothermometry techniques can be compared using the relative sensitivity, defined as \cite{rai2007temperature}
\begin{equation}
S_r=\frac{dX/dT}{X},
\end{equation} 
which allows for standardizing the various methods regardless of the difference in underlying working principle and measured observable, $X$. According to this definition, the sensitivity is expressed in \% K$^{-1}$ units. Typical values of the relative sensitivity are on the level of 0.1--10~\% K$^{-1}$ for PL-based nanothermometers around room-temperature. In turn, Raman-based techniques exhibit S values around 0.01--0.1~\% K$^{-1}$ as shown in Figure~\ref{Thermo_chart}.
 
\textbf{Temperature range, robustness, and stability.}
One of the main shortcomings of thermosensing techniques based on organic dyes is their limited operating range -- a few tens of degrees around room temperature. For example, dye molecules integrated with a resonant silicon dimer were burned at relatively low temperatures.~\cite{caldarola2015non} Record-high temperature range up to $\Delta$T=750~K for PL-based nanothermometers was demonstrated in a specific nanomaterial based on oxide NPs doped with rare earth and covered by Au NPs~\cite{debasu2013all}. In turn, silicon nanoparticles were successfully tested in the range of T=300--1000~K~\cite{zograf2017resonant,aouassa2017temperature, milichko2018metal, karpinski2020optical, odebo2020optical}, where any thermally-sensitive organics would die or PL would vanish because of exponentially increased nonradiative losses. Also, Raman-emitting approaches are more stable and robust in general, because they are not affected by any additional FRET-based or oxygen-based channels of nonradiative recombination during the measurement process. 

\begin{table}[]
\resizebox{\textwidth}{!}{%
\begin{tabular}{l|l|l|l}
    & Materials for nanothermometry             & S$_{r}$, {[}K$^{-1}${]} at 300 K   & $\Delta$T{[K]}  \\ \hline
Organic dyes (ODs)   & Ruphen (PL intensity)~\cite{kose2005preparation}           & 0.0093     & 280-315  \\
      & Bis (pyrene) propane (PL ratio)~\cite{migler1998fluorescence} & 0.387            & 310-465  \\
      & Rhodamine-B (PL Intensity)~\cite{ross2001temperature}      & 0.02             & 287-363  \\
      & Fluorescein (PL anisotropy)~\cite{baffou2009temperature}     & 0.057             & 293-352  \\
      & Triarylboron (wavelength shift)~\cite{feng2011triarylboron} & 0.005            & 223-373  \\ \hline
Quantum dots (QDs) & CdSe (wavelength shift) ~\cite{li2007single}     & 1.61$\times$10$^{-4}$ & 293-323          \\
      & CdSe/ZnS (PL intensity)  ~\cite{walker2003quantum}      & 0.019              & 278-313  \\
    & CdSe/Zns (wavelength shift)~\cite{jorge2007optical} & 2.67 $\times$10$^{-4}$     & 284-320          \\
      & Zn$_{1-x}$Mn$_x$Se/ZnCdSe (PL ratio) ~\cite{vlaskin2010tunable}   & 0.018     & 134-400  \\
      & CdTe (PL lifetime) ~\cite{haro2012high}            & 0.008      & 293-333  \\ \hline
Up-conversion nanoparticles (UCNPs) & Er$^{3+}$/Yb$^{3+}$ CaF$_2$ (PL ratio) ~\cite{dong2011nir}  & 0.015    & 293-318  \\
      & Tm$^{3+}$/Yb$^{3+}$ CaF$_2$ (PL ratio)~\cite{dong2011nir}      & 0.002           & 293-318  \\
      & NaYF$_4$:Er$^{3+}$, Yb$^{3+}$ (PL ratio)~\cite{vetrone2010temperature}     & 0.0114       & 298-334  \\
      & NaLuF$_4$:Yb, Er~\cite{zhu2016temperature}                   & 0.009            & 273-348  \\
      & ZnO:Er$^{3+}$ (PL ratio)~\cite{wang2007effect}            & 0.0098          & 278-463  \\
      & $\beta$-NaYF$_4$:20$\%$Yb$_2\%$Er (PL ratio)~\cite{green2018optical}  & 0.0157       & 294-334  \\
      & (Gd,Yb,Er)$_2$O$_3$ (PL ratio) ~\cite{debasu2013all}     & 0.017       & 300-1050 \\ \hline
Nanodiamonds (NDs)   & NV (PL intensity)~\cite{plakhotnik2015all}               & 0.01       & 295-400  \\
      & GeV (ZPL linewidth)~\cite{fan2018germanium}           & 0.0064    & 150-400  \\
      & SiV (ZPL shift)~\cite{nguyen2018all}                 & 1.61 $\times$10$^{-5}$ & 285-305  \\
      & SnV (ZPL shift)~\cite{alkahtani2018tin}    & 8.66 $\times$10$^{-5}$ & 295-315  \\
      & GeV (anti-Stokes)~\cite{tran2019anti}             & 0.014  & 150-400 \\ \hline 
Optically resonant nanoparticles (RNPs)  & Au nanorod (anti-Stokes PL)~\cite{carattino2018gold}     &   10$^{-3}$    &  300-1300~\cite{buffat1976size}\\ 
      & Si (Stokes Raman shift)~\cite{zograf2017resonant}     &  2$\times$10$^{-4}$    &  0-1685~\cite{nemanich1984raman,talyzin2019size, scheel2008silicon}\\ 
      & $\alpha$-Fe$_2$O$_3$ (Stokes Raman shift)~\cite{zograf2019all}     & 4$\times$10$^{-4}$ &  0-1840~\cite{lide2000handbook}\\ 
\end{tabular}%
}
\caption{Nanothermometers. List of materials and nanothermometry techniques based on light-emitting materials, nanostructures, and nanomaterials. Adopted from~\cite{bradac2020optical} and extended by resonant plasmonic and non-plasmonic nanoparticles.}
\label{tab:my-table}
\end{table}

\section{Applications of all-dielectric thermonanophotonics}
\begin{figure}
\centering
  \includegraphics[width=.99\textwidth]{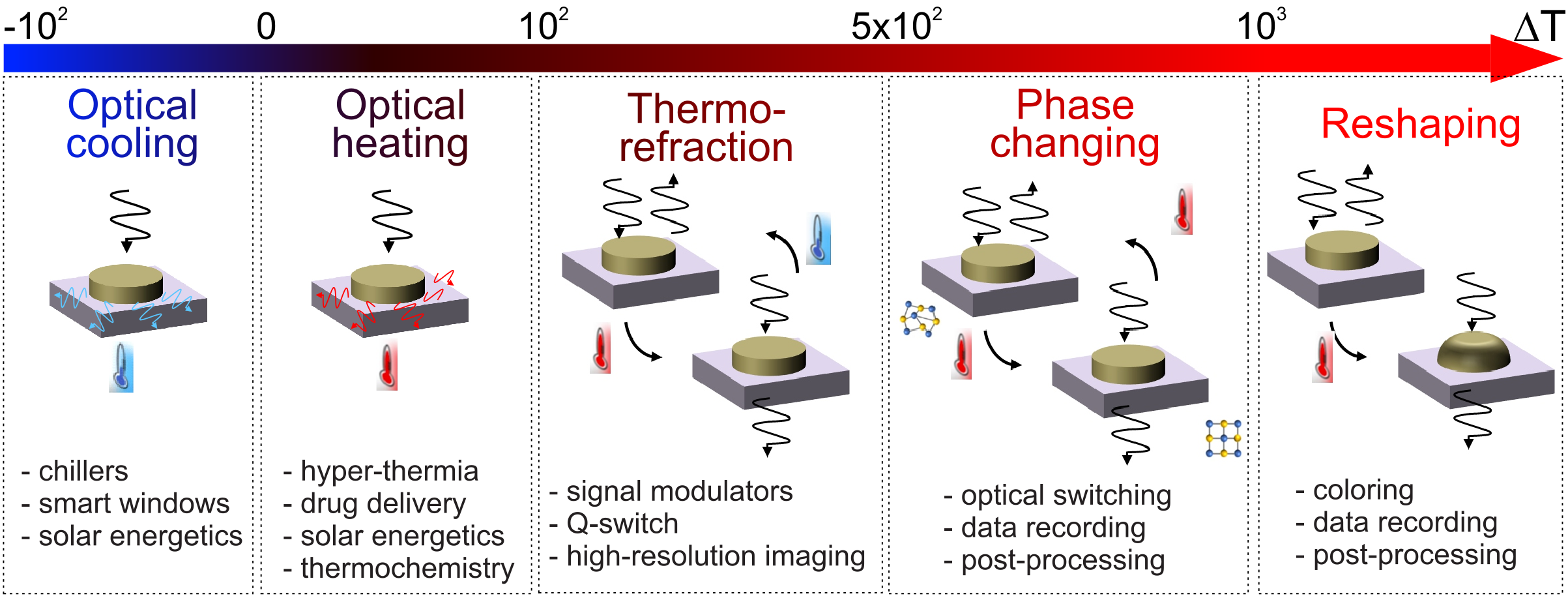}
  \caption{\textbf{Applications of thermally-induced processes.} Conceptual images of different thermally-induced processes presented on the temperature scale,  
  from lower to higher. 
}
  \label{fig:applications}
\end{figure}

The described in previous sections physical phenomena can be employed in variety of applications which are illustrated in Fig.~\ref{fig:applications}. The applications will appear according to the enhancement of temperature from low to high, which is generally correlated with the range of light sources irradiating the nanostructures: from broadband solar light to highly intensive laser radiation.

\subsection{Basic properties of materials employed for thermonanophotonics}

The main physical parameters of bulk dielectric materials shown in Table 2 play key role in the relation to real applications of thermo-photonics. Indeed, according to the above mentioned theoretical models, the knowledge of thermal conductivity, thermal capacity and band gap determine the heating efficiency in the given spectral range, while thermo-optical coefficient and thermal expansion coefficient are important for additional nonlinear feedback of the heated nanostructure.


\begin{table}[]
\resizebox{\textwidth}{!}{%
\begin{tabular}{|c|c|c|c|c|c|c|}
\hline
\multirow{2}{*}{\begin{tabular}[c]{@{}c@{}}Material\\ (Crystalline state)\end{tabular}} &
  \begin{tabular}[c]{@{}c@{}}Melting\\ Temperature\end{tabular} &
  \begin{tabular}[c]{@{}c@{}}Heat \\ Capacity\end{tabular} &
  \begin{tabular}[c]{@{}c@{}}Thermal\\ Conductivity\end{tabular} &
  \begin{tabular}[c]{@{}c@{}}Thermo-optical\\ coefficient\end{tabular} &
  \begin{tabular}[c]{@{}c@{}}Thermal expansion\\ coefficient\end{tabular} &
  Band gap \\ \cline{2-7} 
 &
  K &
  J/(g K) &
  W/(m K) &
  $10^{-6}~\textrm{K}^{-1}$ &
  $10^{-6}~\textrm{K}^{-1}$ &
  eV \\ \hline
Si (c) &
    1687~\cite{adachi2009properties} &
  0.713~\cite{adachi2009properties} &
  156~\cite{adachi2009properties} &
  \begin{tabular}[c]{@{}c@{}}250(\textgreater{}2 $\mu \textrm{m}$)\\ 150 (\textgreater{}2$\mu \textrm{m}$)\end{tabular}~\cite{ghosh1998handbook} &
  2.616~\cite{adachi2009properties} &
  1.14(i)~\cite{levinshtein1997handbook} \\ \hline
Si (a) &
  1400~\cite{tsang1993calorimetric} &
  21 J/mol/K~\cite{tsang1993calorimetric} &
  1.8~\cite{zink2006thermal} &
  \begin{tabular}[c]{@{}c@{}}-147+485i($@$633 nm)\\ 271+225i($@$752 nm)\\ \red{+imaginary part}\end{tabular}~\cite{yavas1993temperature} &
  $\sim$1~\cite{takimoto2002linear} &
  1.14(i)~\cite{levinshtein1997handbook} \\ \hline
Ge &
  1210.4~\cite{adachi2009properties} &
  0.3295~\cite{adachi2009properties} &
  60~\cite{adachi2009properties} &
  \begin{tabular}[c]{@{}c@{}}400(\textgreater{}2 $\mu \textrm{m}$)\\ 400-500 (\textgreater{}2$\mu \textrm{m}$)\end{tabular}~\cite{ghosh1998handbook} &
  5.75~\cite{adachi2009properties} &
  0.67(i)~\cite{levinshtein1997handbook} \\ \hline
GaAs &
  1513~\cite{adachi2009properties} &
  0.327~\cite{adachi2009properties} &
  45~\cite{adachi2009properties} &
  250 ($@$1.15 $\mu \textrm{m}$)~\cite{ghosh1998handbook}&
  6.03~\cite{adachi2009properties} &
  1.43(d)~\cite{levinshtein1997handbook} \\ \hline
GaP &
  1730~\cite{adachi2009properties} &
  0.313~\cite{adachi2009properties} &
  77~\cite{adachi2009properties} &
  160 ($@$0.63 $\mu \textrm{m}$)~\cite{ghosh1998handbook}&
  4.89~\cite{adachi2009properties} &
  2.26(d)~\cite{levinshtein1997handbook} \\ \hline
CdTe &
  1365~\cite{adachi2009properties} &
  0.211~\cite{adachi2009properties} &
  7.5~\cite{adachi2009properties} &
  147($@$1.15 $\mu \textrm{m}$)~\cite{ghosh1998handbook}&
  4.7~\cite{adachi2009properties} &
  1.49(d)~\cite{levinshtein1997handbook} \\ \hline
PbTe &
  1197 &
  0.0031~\cite{el1983thermophysical} &
  1.98~\cite{el1983thermophysical} &
  -1.4$\times 10^{-3}$ (\textgreater{}6$\mu \textrm{m}$)~\cite{lewi2019thermally} &
  19.8~\cite{skelton2014thermal} &
  0.32(d)~\cite{skelton2014thermal}\\ \hline
Ge$_2$Sb$_2$Te$_5$ (a,c) &
  \begin{tabular}[c]{@{}c@{}}900\\ (a-\textgreater{}c $\sim$400 )\end{tabular}~\cite{balde1995etude,morales2005structural} &
  0.212~\cite{battaglia2010thermal} &
  \begin{tabular}[c]{@{}c@{}}0.19,\\ 0.57-1.5\end{tabular}~\cite{lyeo2006thermal} &
  \begin{tabular}[c]{@{}c@{}}a: $\Delta$k = 11.7 \\ $\Delta$n = 35\\ c: $\Delta$k = 113\\ $\Delta$n = -65 ($@$1.55 $\mu \textrm{m}$)\end{tabular}~\cite{stegmaier2016thermo} &
  2 -- 7 ~\cite{jong2001mechanical}&
  0.95(d)~\cite{levinshtein1997handbook} \\ \hline
MAPbCl$_{3}$ &
  650 ~\cite{brunetti2016thermal} &
  0.492~\cite{handa2019large} &
  0.46~\cite{handa2019large} &
  -300~\cite{handa2019large} &
  30~\cite{ge2018ultralow} &
  3.15~\cite{handa2019large} \\ \hline
Au &
  1337~\cite{baffou2013thermo} &
  0.129~\cite{baffou2013thermo} &
  318~\cite{baffou2013thermo} &
  -300 -- -700~\cite{sarkhosh2010large} &
  14~\cite{nix1941thermal} &
  \multirow{3}{*}{Metals} \\ \cline{1-6}
Ag &
  1235~\cite{baffou2013thermo} &
  0.24~\cite{baffou2013thermo} &
  429~\cite{baffou2013thermo} &
  -300 -- -700~\cite{karimzadeh2010effect} &
  18.9~\cite{nix1941thermal} &
   \\ \cline{1-6}
Al &
  933~\cite{baffou2013thermo} &
  0.9~\cite{baffou2013thermo} &
  237~\cite{baffou2013thermo} &
  \multicolumn{1}{l|}{C$_{TR} = 114$ ($@$780 nm)~\cite{favaloro2015characterization}} &
  23.3~\cite{nix1941thermal} &
   \\ \hline
\end{tabular}}
\caption{Properties of the materials usually used for thermophotonic applications. Parameters are taken at ambient conditions if not noted otherwise. Indices in brackets for band gap denote direct (d) or indirect (i) band. Indices in brackets for material names are denoted crystalline phase - amorphous (a) or crystalline (c). C$_{TR}$ stands for thermoreflectance coefficient.}
\label{tab:my-table}
\end{table}



\subsection{Optical cooling}

\begin{figure}
\centering
  \includegraphics[width=.85\columnwidth]{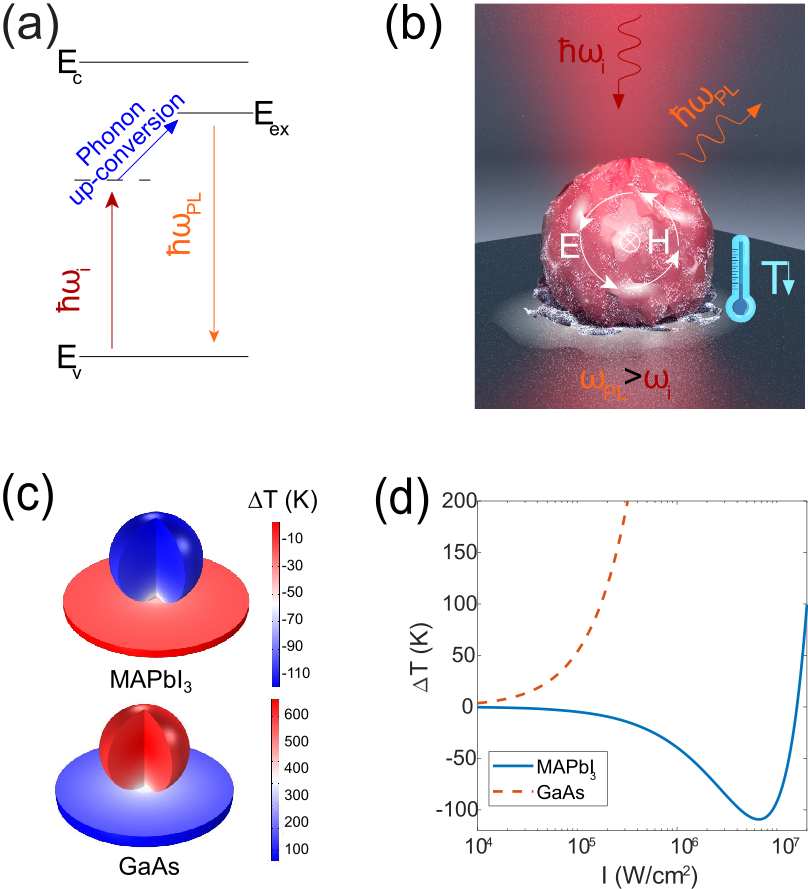}
  \caption{\textbf{Optical cooling.} (a) Scheme of photoluminescence upconversion mechanism. (b) Sketch of optical cooling via enhanced, upon excitation of Mie-modes, upconversion photoluminescence. (c) Temperature distributions for 530~nm \(\mathrm{MAPbI_3}\) NP under 980~nm laser illumination with \(\mathrm{7\cdot10^{6}W/cm^2}\) intensity and 340~nm GaAs NP under laser illumination with the same wavelength and  \(\mathrm{10^{6}W/cm^2}\) intensity. (d) The dependence of temperature change for 530~nm \(\mathrm{MAPbI_3}\) NP (blue solid line) and 340~nm GaAs NP (red dashed line) on substrate on laser intensity at wavelength $\lambda$=980~nm.~\cite{tonkaev2019optical}}
  \label{fig:cooling}
\end{figure}

In 1929, it was suggested that solids could cool through anti-Stokes fluorescence in which a substance absorbs a photon and then emits one of greater energy~\cite{pringsheim1929zwei} (so-called up-conversion) via absorption of an additional phonon for energy conservation (see Fig.~\ref{fig:cooling}a).

For decades the advantages of laser cooling of solids were connected with RE-doped glasses and crystals. In 1950, it was proposed the use of rare-earth ions in transparent solids as a fluorescent cooler because of their high quantum efficiencies and narrow spectral lines~\cite{kastler1950some}. The main advantage of RE ions is the optically active \textit{4f} electrons shielded by the filled 5s and \textit{5p} outer shells, which limit interaction with the lattice surrounding the RE ion and suppress nonradiative decay. The successful realization of laser cooling of rare-earth-doped solids was demonstrated in 1995~\cite{epstein1995observation}. Since then, laser-induced cooling has been observed in a wide variety of glasses and crystals doped with ytterbium (Yb$^{3+}$), thulium (Tm$^{3+}$) and erbium (Er$^{3+}$).~\cite{sheik2009laser, nemova2010laser} Among the various dielectric materials, low trap-state density, high PL quantum yield, as well as pronounced excitonic states at room temperature in bulk halide perovskites resulted in high efficiency of up-conversion, and, thus, in the decrease of local temperature by 20~K upon laser irradiation~\cite{ha2016laser}.

There are two ways to enhance optical cooling as it follows from Eq.~\ref{Nc}. Firstly, one can increase external radiative quantum efficiency via decreasing the radiation lifetime. Secondly, one can decrease frequency of absorbed light. Balance of these two effects determines the cooling efficiency, which can be expressed by the ratio introduced in Ref.~\cite{sheik2004can} as following:  
\begin{equation}\label{Nc}
\eta_c=\frac{P_{PL}}{P_{PL}+P_{abs}}=\eta\frac{\omega_{PL}}{\omega_i}-1~,
\end{equation}
where $\eta$ is luminescence quantum efficiency, $\omega_{PL}$ is PL frequency, $P_{PL}(P_{abs})$ is the emitted (absorbed) power of light.

In order to determine the temperature change of the NP under laser illumination, one should consider the stationary heat transfer equation and take into account only radial heat distribution: 
\begin{equation}\label{cool}
\nabla(\kappa\nabla T)=\eta_c\frac{I\sigma_{abs}}{V}~,
\end{equation}
where $\kappa$ is thermal conductivity.

Further optimization of the optical cooling requires both enhanced quantum yield of emission and improved absorption of incident light in the material.
Recently, it was proposed that the resonant phonon-assisted up-conversion photoluminescence optical cooling approach can be optimized via enhancement of the emission rate and photoexcitation with Mie resonances in nanoparticles at pump and emission wavelengths (see schematic illustration in Fig.~\ref{fig:cooling}b).~\cite{tonkaev2019optical} 
In this case, the expression for the temperature variance inside the NP can be found by solving the equation for thermal diffusion and appears to be following:
\begin{equation}\label{Eq:cooling}
\Delta T= -\eta_c \frac{\sigma_{abs}I}{2\pi\kappa D}~,
\end{equation}
where $\kappa$ is the thermal conductivity of the surrounding medium and $D$ is the diameter of a nanoscale sphere. Remarkably, that $\eta_c$ is directly connected with Purcell effect and, thus, cooling efficiency can be enhanced when emission spectra is overlapped with a resonant mode in the nanoparticle. 

The numerical and analytical modeling revealed that the highest cooling efficiencies for a halide perovskite spherical NP correspond to the excitation of magnetic-type Mie modes.~\cite{tonkaev2019optical} Namely, magnetic octupole at the emission and magnetic quadrupole at absorption allow for cooling a single nanocavity by $\Delta T$~$\approx$~-110~K at realistic conditions. In opposite, GaAs nanoparticles with less efficient PL can not be optically cooled with this mechanism as shown in Fig.~\ref{fig:cooling}c,d. 

\subsection{Solar energy and heat conversion}

In our everyday life, the lowest irradiation intensities capable to heat various objects are coming from the Sun, i.e. around 100~W/m$^2$. A typical sunlight application is a solar water heater placed usually on a roof to collect solar energy. Meanwhile, many other applications recently appeared for photothermal devices based on solar energy: water desalination, surface sterilization, deicing, water evaporators, etc.~\cite{wu2021solar,chenphotothermal}. One of the typical design is to use separate sunlight absorbers (some black surfaces), which convert solar energy into thermal energy that is then transferred to water.
However, a more efficient way of water heating is heating through nanoparticles dispersed in water, which is called nanofluid~\cite{otanicar2009comparative}. In this case, absorbing and generating heat nanoparticles transfers it directly to the water. Thus, the main challenge here is to provide efficient sunlight harvesting in as broad spectral range as possible and high light-to-heat conversion efficiency.

\begin{figure}
\centering
  \includegraphics[width=.75\columnwidth]{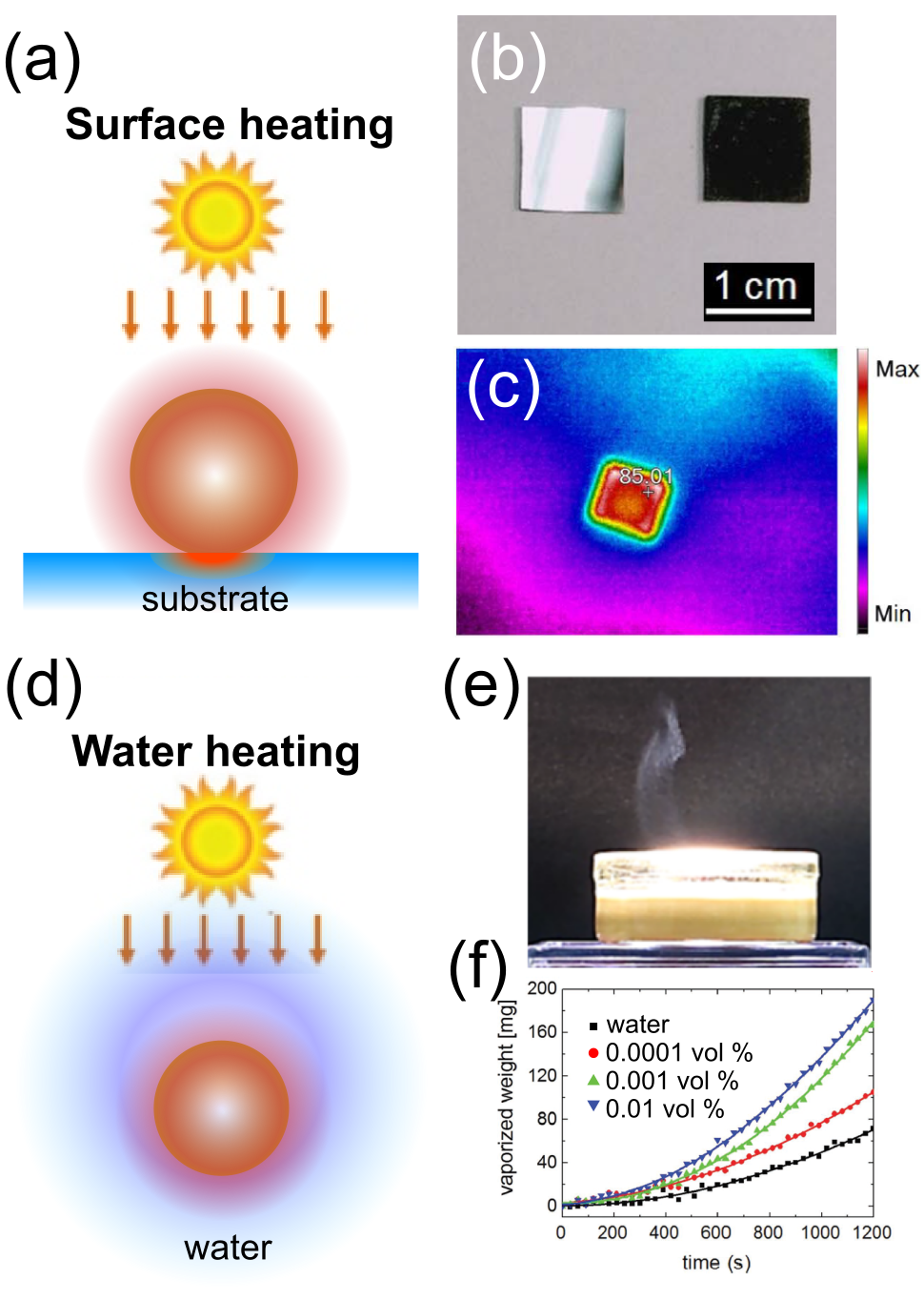}
  \caption{\textbf{Solar energy applications.} (a) Schematic diagram of a nanoparticle on surface irradiated by sunlight. (b) Photograph of a bare Si wafer (left) and Te nanoparticle layer deposited on Si substrate (right). (c) Steady-state thermal image of the Te nanoparticle absorber.~\cite{ma2018optical} (d) Schematic diagram of a nanoparticle in water irradiated by sunlight. (e) The photo of 0.1 vol\% Si nanofluid at the irradiation of~800 mW/cm$^{2}$ from the solar simulator~\cite{ishii2016solar} (e) Vaporized weight of the water under the illumination of simulated sunlight at 80 mW/cm$^{2}$. The concentration of Si nanoparticles in nanofluid is varied from 0 to 0.01~vol\% where 0 vol\% means pure water~\cite{ishii2017resonant}.}
  \label{fig:solar}
\end{figure}

Resonant nanoparticles made of semiconductor with indirect band gap absorption (like Si~\cite{ishii2016solar}, Ge~\cite{ishii2017resonant}, and Te~\cite{ma2018optical}) are the best candidates for such spectrally broadband light energy conversion to heat. Indeed sunlight on the Earth surface is quite spectrally broadband ($\lambda\approx$300--2000~nm) source of energy. An example of water evaporation upon sunlight illumination is shown in Fig.~\ref{fig:solar}e.
Figure~\ref{fig:solar}f depicts the evolution of water evaporation at different Si nanoparticles concentrations~\cite{ishii2016solar}. The ability of Si NPs for broadband light absorption with subsequent conversion to thermal energy by means of Mie-resonances results in high light energy conversion efficiency of the nanofluid (up to 55\%), which is the sum of energies consumed to vaporize $E_v$ water and heat $E_h$ water divided to the energy of incident light $E_{in}$:
\begin{equation}
\eta_{heat} = \frac{E_v + E_h}{E_{in}}
\end{equation}

The light energy conversion efficiencies for various types of nanofluids based on Mie-resonant NPs are summarized in Table~\ref{tab:Solar}

\begin{table}[h!]
\centering
\begin{tabular}{|l|l|l|}
\hline
nanofluid type & $\eta_{heat}$ & ref.                                      \\ \hline
water          & 29 \%                         & \cite{ishii2016solar}    \\ \hline
Si NPs@water   & 55 \%                         & \cite{ishii2016solar}    \\ \hline
Ge NPs@water   & 67 \%                         & \cite{ishii2017resonant} \\ \hline
Te NPs@water   & 85 \%                         & \cite{ma2018optical}     \\ \hline
\end{tabular}
\caption{Energy conversion efficiencies for various types of nanofluids upon sunlight illumination.}
\label{tab:Solar}
\end{table}

Generally, Si and Ge NPs are non-toxic materials, and can be used in liquids for many of real-life applications. Although tellurium NPs were also utilized for efficient water heating upon sunlight illumination~\cite{ma2018optical}, tellurium is considered to be toxic. However, it plays an important role in many biological systems, while the content of tellurium in the human body is more than 0.5~g, being the fourth most abundant trace element after Fe, Zn, and Rb in the human body and it is unusually abundant in human food and plants~\cite{cohen1984anomalous}. Further progress in this field can be done with hybrid metal-dielectric NPs, where Mie-like and plasmonic resonances coupling results in high light energy conversion efficiencies (e.g. with TiO$_2$/Au NPs ~\cite{gurbatov2020decorated}).

\subsection{Biomedical applications}

All-dielectric resonant nanostructures are widely used for optical sensing of various bio-objects like proteins, antibodies, and related molecules~\cite{bosio2019plasmonic, tittl2018imaging}, where the role of Mie resonances is crucial~\cite{yavas2019unravelling}. Despite thermal effects are ignored in the most of works on this topic, temperature growth can be considerable if the nanostructures absorb high enough power of the incident light.   

According to Sections 3 and 4, once local temperature can be measured directly, two main regimes of light interaction with all-dielectric nanoresonators can be distinguished. Fig.~\ref{fig:Hotspots} shows experimental data for the cases when two optically coupled silicon nanoparticles are not heated upon high-intense laser illumination (Fig.~\ref{fig:Hotspots}a), and strongly heated a single resonant silicon nanoparticle at much lower intensity (Fig.~\ref{fig:Hotspots}b). The reason of such big difference lies in the amount of light energy accumulated inside and outside the nanoresonators. Indeed, the silicon nanodimer shown in the work~\cite{caldarola2015non} is an optimal design for PL and SERS in the cold regime when parasitic heating is diminished, because hot spot is concentrated in the gap between nanoparticles, while much less energy accumulated in the silicon parts. However, once nanoparticles are coupled with light resonantly, the opposite situation can be observed when the nanoparticle undergoes strong heating, which was shown experimentally~\cite{zograf2017resonant}. 

Similar transition between `cold' and `hot' regimes of interaction was realized for a silicon nanoparticle on gold~\cite{milichko2018metal}, behaving like a dimer because of formation of `mirror image' modes~\cite{xifre2012mirror}. Moreover, this design is prospective for sensing applications~\cite{hutter2013near, ding2016nanostructure} owing to high near-field concentration in the gap between a nanoparticle and a metal. First SERS measurements in the gap between a resonant dielectric nanoparticle and metal was carried out by Huang et al.~\cite{huang2015strong} and showed up to 10$^7$ Raman signal enhancement. The advantage of existence of thermally sensitive part in this design is in the ability to study \textit{in situ} local Raman response of the material (vibrational strength and spectral changes) in the gap between nanoparticle and metal with simultaneous temperature control in broad range~\cite{milichko2018metal}. 
Controllable local heating and modification of surrounding material with all-dielectric nanoantennas was also employed for intracellular optical opening of microcapsules for drug delivery applications~\cite{zograf2019all}.

Despite the ability of dielectric resonant nanostructures to be heated efficiently by light in the visible range, biological applications related to IR sources are not affected by any thermal effects, and can be safely employed~\cite{tittl2018imaging, yesilkoy2019ultrasensitive}.

\begin{figure}
\centering
  \includegraphics[width=.95\textwidth]{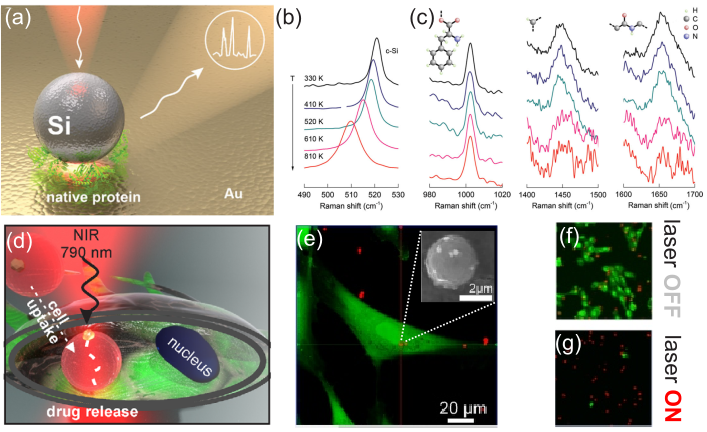}
  \caption{%
\textbf{Biomedical applications.} (a) Schematic illustration of the hybrid nanocavity for simultaneous molecular sensing, nanoscale thermometry, and tracing the heat‐induced events through the change in Raman scattering signal. (b) Heat‐induced evolution of normalized Raman spectra for 190~nm c‐Si NP of the nanocavity and (c) BSA molecules inside. 
(d) Schematic of cancer cell uptake of polymer capsules loaded with anti-tumour drug with subsequent thermally triggered release via heating of $\alpha$-Fe$_2$O$_3$ NPs embedded into capsules walls with real-time temperature control by Raman scattering. (e) Confocal laser
scanning microscopy images of the polymer capsules with $\alpha$-Fe$_2$O$_3$ nanoparticles in living cells. Inset shows SEM image of the capsule with $\alpha$-Fe$_2$O$_3$ nanoparticles. Cancer cells viability after incubation with polymer capsules modified with $\alpha$-Fe$_2$O$_3$ NPs and loaded with antitumor drug VCR (vincristine) for 24 h before (f) and after (g) irradiation with NIR laser.}
  \label{fig:drug}
\end{figure}

\subsection{Thermorefractive optical nonlinearities}

\begin{figure}
\centering
  \includegraphics[width=.75\textwidth]{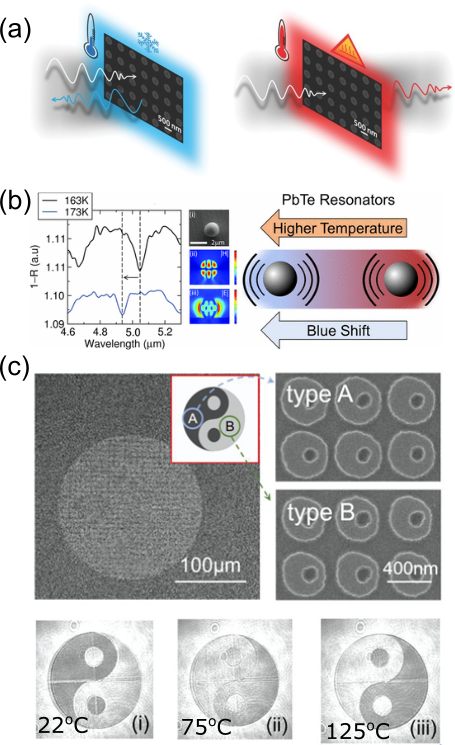}
  \caption{%
\textbf{Thermal tuning of metaphotonic devices.} (a) Illustration of the temperature effect on the far-field optical properties of metasurfaces by employing the actual scanning electron microscope (SEM) image. By cooling or heating the metasurface sample, one can operate in either the reflection or transmission regime, respectively~\cite{rahmani2017reversible}. (b) Ultrawide spectral tuning of resonances via heating of semiconductor meta-atoms~\cite{lewi2017ultrawide}. (c) The SEM images of the fabricated Yin-Yang pattern with slightly different geometries of the two parts. The optical images of the metasurfaces obtained at 784 nm at different temperatures~\cite{zangeneh2019reversible}.}
  \label{fig:Thermo_tune}
\end{figure}

With the increase of the temperature the optical constants of the solids can not be generally considered as constant and start to depend on temperature, which is referred to as {\it thermo-optical effect}. There are  several physical mechanisms lying in the origin of this effect. One of the most important is related to the dependence of the electronic band gap on the temperature \cite{ashcroft1976nd} due to  thermal expansion of lattice. That, along with the temperature dependent Fermi level, gives the most significant contribution  in thermo-optical effect in semiconductor materials. Another contribution is related to the increased  scattering rate of electrons in solids due to the enhanced phonon scattering, however this mechanism in more important in metals where the electron concentration stays almost constant. 

The thermally induced refractive index change has real and imaginary parts being linearly dependent on temperature given by the following expression:
\begin{equation}
\Delta n~=~\Gamma (T-T_0), \quad \Gamma=\left.\dfrac{dn}{dT}\right|_{T_0}
\end{equation}
where $\Gamma$ is the complex thermo-optical coefficient. Silicon has relatively high $\Gamma$, the real part of the $\Gamma$ is equal to 4.5$\cdot$10$^{-4}$~K$^{-1}$ and the imaginary part is equal to 0.1$\cdot$10$^{-4}$~K$^{-1}$~\cite{esser1989ultrafast}, whereas GaAs has four times smaller corresponding values. The values of the thermo-optical coefficients are summarized in Table 1. One can notice that the most of the semiconductor materials have positive real part of $\Gamma$. Thermal expansion leads to longer interatomic distances and, thus,  weaker interaction between the electronic states and consequent decrease in the band gap. The well-known general law describing the temperature dependence of the band gap is as follows:  

\begin{equation}
E_g(T)=E_g(0)-\dfrac{\xi_1 T^2}{T+\xi_2},
\end{equation}
where $\xi_1$ and $\xi_2$ are material dependent constants. The simple reasons based on thermal expansion provide that for the majority of materials $\xi_1>0$, which results in positive thermo-optical coefficient $\Gamma'>0$. However,  there is a number of materials where the band gap temperature dependence has anomalous form and, for instance, $\xi_1<0$ resulting in $\Gamma'<0$. One can see that PbTe is one of that type of materials with negative  of thermo-optical constant being several times higher in absolute values\cite{lewi2017ultrawide,lewi2018thermal}, comparing to others materials. Recently,  high  negative thermo-optical constant of lead halide perovskites was also reported \cite{handa2019large}.    

From the point of resonant nanoscale heating structure, the temperature change of the refractive index leads to thermal drift of the resonant wavelength: 

\begin{equation}
\dfrac{\Delta \lambda_0}{\lambda_0}\sim \left(\dfrac{1}{n}\dfrac{dn}{dT}  + \dfrac{1}{D} \dfrac{dD}{dT}\right)\Delta T.
\end{equation}
Here, the first term corresponds to thermo-optical coefficient, while the second one is related to thermal expansion coefficient and related change in geometrical size of the nanostructure. From Table 1 one can see that normally the thermal expansion coefficient is smaller than the thermo-optical constant, but the contributions to the relative wavelength shift may be comparable. The recent findings show that among other materials perovskite structures may posses quite high value of thermal expansion coefficient $250\cdot10^{-6}$ K$^{-1}$\cite{ge2018ultralow}.

Remarkably, typical raising time of the thermal nonlinearity is governed by electron-phonon scattering timescale laying in the interval of 1--10 ps~\cite{downer1986ultrafast}. The relaxation time in this case is slower than for optically induced nonlinearities caused by Kerr effects~\cite{grinblat2020efficient} or free carriers~\cite{makarov2015tuning, shcherbakov2015ultrafast} generation being usually about 1--100~ns~\cite{duh2020giant} strongly depending on the thermal conductivity of surrounding medium.

In the case of a single resonant nanoparticle, it was shown theoretically that the photo-induced thermo-optical effect can lead to a self-consistent nonlinear heating~\cite{tsoulos2020self}.

Thermo-refractive optical nonlinearity was successfully utilized in various applications related to dramatic reversible changes of the optical properties of nanophotonic designs during temperature modulation.

In work~\cite{lewi2017ultrawide}, optical properties of microspheres from PbTe ($\Gamma\approx$--15$\cdot$10$^{-4}$~K$^{-1}$)
materials with Mie resonances were tuned in infrared range by varying temperature in range of 80--573~K~\ref{fig:Thermo_tune}b. This approach allowed to reversibly reconfigure all-dielectric nanoantennas resonances over their full-width, providing strong modulation of scattered/transmitted optical signal. In particular, it was demonstrated that high-quality factor Mie resonances can be tuned by several linewidths with temperature modulations as small as $\Delta$T$\sim$~10~K.

Next step was done for all-dielectric metasurfaces, where the temperature-dependent change of the refractive
index of silicon was employed to tune light transmission (see scheme in Fig.~\ref{fig:Thermo_tune}a) in a spectral window of 75~nm around the telecom wavelength.~\cite{rahmani2017reversible} The heating process resulted in a significant changes in the forward to backward light power propagation ratio from around 1 to more than 50 times. This is one of the highest reported value for tuning directionality by means of a reversible technique.

Basing on this approach, reversible image tuning was demonstrated with temperature tunable dielectric metasurfaces.~\cite{zangeneh2019reversible} A metasurface with the encoded transmission Yin-Yang pattern (see Fig.~\ref{fig:Thermo_tune}c) was heated up by just 100~K. Thermo-refractive nonlinear optical modulation of silicon properties was strong enough to change spectrally sharp Fano resonances of two sets of nanoresonators composed of nonconcentric silicon disks with holes (see SEM images in Fig.~\ref{fig:Thermo_tune}c). The authors achieved full control of the contrast of the Yin-Yang image in the reversible manner.
Such thermally sensitive designs might be prospective for power limiters~\cite{hsu2021broadband}.  


\begin{figure}
\centering
  \includegraphics[width=.79\textwidth]{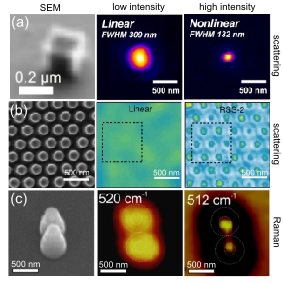}
  \caption{%
 \textbf{Superresolution imaging of Si nanoparticles with thermo-optical nonlinearity.} (a) SEM image of a Si nanocube on a quartz substrate, and optical images of scattered signal from a nanoscale cube at low (linear regime) and high (nonlinear regime, $\sim$6~mW/$\mu$m$^2$ at wavelength 592~nm) intensities of incident CW laser.~\cite{duh2020giant} (b) Confocal optical images of nonlinear scattering from periodic Si nanodisk arrays with anapole states evolve with increasing excitation intensities and correlated SEM image. The optical super-resolution image is a differential between two images, where the nonlinear regime corresponds to 15~mW/$\mu$m$^2$ at wavelength 532~nm.~\cite{zhang2020anapole} (c) Side-view SEM images of the Si pillar dimer, as well as corresponding Raman scattering maps at 520 cm$^{-1}$ and 512 cm$^{-1}$ measured at the incident intensities of 4 and 42~mW/$\mu$m$^2$ at wavelength 532~nm, respectively.~\cite{aouassa2017temperature}}
  \label{fig:imaging}
\end{figure}


Stronger laser-induced heating of silicon allowed more than +400\% to -90\% nonlinear deviation of scattering from a single Mie-resonant silicon nanoparticle under continuous-wave (CW) illumination via enhanced thermo-refractive nonlinearity.~\cite{duh2020giant} The nanostructure can exhibit a strong nonlinear response (saturation of scattering). With a Gaussian focus, the nonlinear response should start from the center of point-source function, and thus, by extracting the nonlinear part, the resulting point-source function becomes smaller than its linear counterpart. It helped to beat diffraction limit by 2.3-times enhancement of optical imaging resolution  enhancement as demonstrated in Fig.~\ref{fig:imaging}a.

In work~\cite{zhang2020anapole}, at the anapole wavelength in Si nanodisk the boosted near-field energy directly contributes to the absorption, leading to a substantial temperature rise within the nanoparticle. In the temperature range from RT to 950~$^o$C, the change of the refractive index in real part $\Delta n$ is extrapolated to be 0.5 at a moderate laser intensity of 1.25$\times$10$^6$~W/cm$^2$. This equivalently gives the effective nonlinear refractive index as $n_2$(532~nm)=$\Delta n$/I=0.4 cm$^2$/MW. Compared with the measured temperature rise in bulk Si, optical anapole significantly enhanced photothermal nonlinearity by three orders of magnitude.

Finally, Raman thermometry can be applied for mapping the non-uniformly heated Si pillars with the sub-diffraction lateral resolution.~\cite{aouassa2017temperature} As shown in Figure~\ref{fig:imaging}c, Raman maps obtained at incident beam intensities of 4 and 42~mW/$\mu$m$^2$ for Si pillar dimer. At I=4~mW/$\mu$m$^2$, which corresponds to almost the “cold” structure, no specific features can be resolved in the acquired Raman map. On the other hand, at I=42~mW/$\mu$m$^2$, the Raman signal distribution at 520~cm$^{-1}$ shows a decrease in the intensity in the central area of each Si pillar, which is attributed to the pronounced temperature dependent spectral shift of the corresponding c-Si band in this particular area upon its more efficient heating. Remarcably, the Raman imaging technique also allows for simultaneous monitoring of the material properties for complex compositions (e.g. like SiGe~\cite{mitsai2019si}). 

All these works on thermal-nonlinearity enhanced optical imaging demonstrated comparable resolution improvement down to $\lambda$/10, extending the application of super-resolution microscopy not only to label-free silicon nanostructure observations but also further into biomedical applications with silicon nanoparticles.


\subsection{Thermally-induced phase transitions}

Further increase of the temperature range leads to inherent phase transitions in the materials associated with a change of material parameters, when the local temperature of a sample exceeds a threshold value. The most of such phase transitions are volatile so that a material returns to its initial phase upon cooling. One of the most common example is melting/crystallization process.

Another class of phase transitions, namely transitions between crystalline and amorphous phases, paves the way for designing a non-volatile photonic nanostructures made of materials with a glass-transition effect. Amorphous phase generally has higher configuration entropy than the lowest free-energy state in crystalline phase. At the same time, below the melting temperature $T_m$ viscosity exhibiting strong increase with the decreasing of the temperature suppresses atomic diffusion at the glass-transition temperature $T_g$. Thus, a rapid quenching of melted material (the cooling rate higher than crystalline rate) prohibits atoms to form an ordered lattice, resulting in formation of a quasi-stable amorphous phase.
In the case of nanophotonic structures, a femtosecond optical pulse that delivers energy to material for heating it above $T_m$ succeeded by quenching the sample below the glass-transition temperature $T_g$ leads to freeze a disorder making the transition non-volatile. However, this process can be reversed. A train of pulses, each of them heats the antenna to the temperature above $T_g$ but below $T_m$, allows transforming the material back into a crystalline phase through a sequence of amorphous phases with decreasing disorder.

\begin{figure}
\centering
  \includegraphics[width=.69\textwidth]{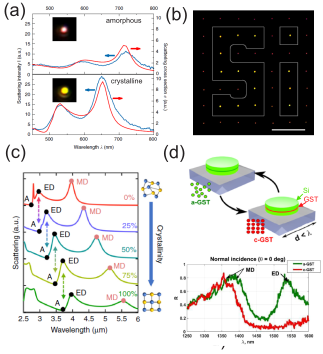}
  \caption{\textbf{Nanoparticles made of phase-changing materials.} (a) Experimental (blue curves) and theoretical (red curves) scattering spectra, calculated using Mie theory, of spherical Si nanoparticles. (b) Dark-field microscopic image of the laser-printed Si nanoparticles. The nanoparticles within the white lines are crystallized by additional laser pulse irradiation, producing a visible colour change (scale bar, 10~$\mu$m)~\cite{zywietz2014laser}. (c) Numerically solved conditions different bright and dark states under different crystallinities of a GST nanodisk~\cite{tian2019active}. (d) hybrid Si/GST cylinders effectively behave as Si-only when the GST is amorphous, and the resonant modes supported by the array (thus its optical response) can be modified on demand by switching the GST layer between its amorphous and crystalline states and experimentally obtained reflectance spectra for the as-fabricated device with the GST layer in both amorphous and crystalline states~\cite{ruiz2020reconfigurable}.}
  \label{fig:thermo-phase}
\end{figure}

Non-volatile phase transitions were demonstrated for silicon nanospheres fabricated by a laser ablation method, i.e. a femtosecond laser printing~\cite{zywietz2014laser, makarov2018resonant}. Amorphous (a-Si) to crystalline (c-Si) transitions in silicon leads to a decrease of dielectric permittivity in the spectral range from 500 nm to 900 nm. The physical origin of this behavior is the following. In c-Si, electron transitions in the interval from 365 to 1130 nm are indirect, that is they take place between states of different wave vector with the simultaneous absorption or emission of a phonon. At the same time the lack of wave vector conservation in a-Si makes these transitions quasi-allowed resulting in stronger optical response . Also there are differences in the density of valence states, which also account for the longer wavelength shift in the maximum of  the imaginary part of $\varepsilon$ \cite{joannopoulos1984physics}. This moderate change of the permittivity (from $\varepsilon_\mathrm{a-Si}=16.5$ to $\varepsilon_\mathrm{c-Si}=14$ at wavelength $\lambda\approx 700$ nm) can shift a position of the Mie resonance up to 70 nm~\cite{zywietz2014laser}. By using the femtosecond laser printing, Zywietz {\em et al.}~\cite{zywietz2014laser} fabricated a square lattice (lattice spacing 5 $\mu$m) of similar a-Si nanoparticles that have the magnetic dipole Mie resonance at $\lambda\approx 720$ nm. Next, the laser-induced crystallization in the array of these a-Si nanoparticles allowed for selective change of the properties of single nanoparticles. After the crystallization, the Mie resonance is shifted to $\lambda\approx 650$ nm since the silicon permittivity is decreased due to the laser-induced phase transition (see Fig.~\ref{fig:thermo-phase}a). This behavior was also employed to modify locally optical properties of individual a-Si nanoparticles to create a picture ``Si'' as shown in Fig.~\ref{fig:thermo-phase}b. 

Further, the approach of local thermally-induced phase switching was applied to record various complicated colorful patterns with amorphous Si nanoparticles prepared by nanolithography.~\cite{wang2018programming} This method is quite useful, because it allows for creation of c-Si nanoparticles on an arbitrary substrate (i.e. like glass) without employing silicon-on-insulator or silicon-on-sapphire technologies, as well as any transfer techniques. Additionally, this technique allowed for local laser annealing of amorphous silicon nanoparticles with \textit{in situ} control of temperature and crystalline state via generated Raman signal analysis~\cite{zograf2018local}.

By using another important compound based on germanium (Ge) -- antimony (Sb) -- tellurium (Te) alloys,  and often referred to as GST, one can also achieve non-volatile transitions. Recently, such GST alloys catch a lot of attention in photonics since their dielectric properties demonstrate a very strong modulation. 
In contrast to the moderate change in optical and electrical properties of amorphous and crystalline phases of \textit{sp}$^3$-bonded semiconductors such as silicon, GST alloys possess unsaturated covalent bonds leading for a resonant bonding to exist in crystalline phase~\cite{lencer2008map}. The resulting ground state can be explained as a superposition of symmetrically-equivalent states with saturated-bond configurations. Thus, electrons are effectively delocalized resulting in a high dielectric permittivity values. On the other hand, the resonant bonding requires the long-range ordering and in amorphous phase this ordering is not possible causing a strong contrast in optical properties of a-GST and c-GST alloys \cite{shportko2008resonant}. 
Due to low losses for $\lambda>1.5~\mu$m the GST-based nanophotonic designs employing optical resonances can operate in the near-infrared range.


Wang {\em et al.}~\cite{wang2016optically} reported on a dipolar metasurface operating around a wavelength of $\lambda=2\mu$m, where absorption in GST is low enough to achieve resonances. They designed the metasurface comprising a two-dimensional array of rectangular crystalline inclusions in the amorphous GST film. Both transmission and reflection spectra show the resonant feature at $\lambda=2\mu$m for light polarized along the inclusion and no dips or peaks for the orthogonal polarization. For wavelengths $\lambda>1.78 \mu$m the structure does not scatter light in the non-zero diffraction orders~\cite{rybin2016transition} demonstrating a true metamaterial nature~\cite{wang2016optically}.

In work~\cite{tian2019active}, it was demonstrated that the structured phase-change alloy Ge$_2$Sb$_2$Te$_5$ (225-GST) can support a diverse set of multipolar Mie resonances with active tunability. By harnessing the dramatic optical contrast of GST,  broadband ($\Delta\lambda/\lambda\sim$15\%) mode shifting between an electric dipole resonance and an anapole state (see Fig.~\ref{fig:thermo-phase}c). Active control of higher-order anapoles and multimodal tuning were also investigated, which make the structured GST serve as a multi-spectral optical switch with high extinction contrasts ($>$6 dB).

In order to shift from IR to visible range, a new concept
in all-dielectric optical metasurfaces introduced and experimentally validated based on a hybrid combination of high-index and low-loss dielectric building blocks with embedded
subwavelength inclusions of chalcogenide phase-change materials.~\cite{ruiz2020reconfigurable}
By using this hybrid approach, the authors were able not only to provide on-demand dynamic control of light amplitude, but also to deliver a very high efficiency of operation over a very wide spectral range by a judicious material choice. The authors demonstrated the flexibility and universality of our approach by the design and development of hybrid metasurfaces for applications as switchable spectral filters in the near-infrared and dynamic color generation in the visible spectrum.

\subsection{Optical reshaping}

\begin{figure}
\centering
  \includegraphics[width=.55\textwidth]{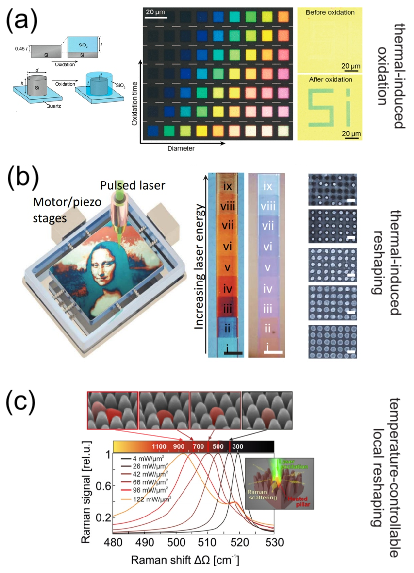}
  \caption{%
(a) Schematic of the calculation model for Si nanostructure oxidation. Reflection image of oxidized arrays through a 20$\times$ objective (NA=0.45) irradiated with linear polarized white light. The diameter was systematically changed from 90~nm to 250~nm in 20~nm increments. The vertical axes show total oxidation time and SiO$_2$ layer thickness. Each individual color area is 10~$\mu$m~$\times$~10 $\mu$m. The black frame corresponds to unstructured regions. The structures shown in each row are the same, with the extent of oxidation increasing toward the top of the image. Scale bar, 10 $\mu$m.~\cite{nagasaki2018control} (b) Schematic setup of resonant laser printing. Synchronous motion solution with the laser pulses is provided by computer-controlled motor or piezostages. Reflection and transmission micro-images of multicolored structures generated by gradually increasing laser powers. Microstructures (i to ix) are generated under gradually increasing laser power strengths from 0.2 to 1.8~mJ in steps of 0.2~mJ, controlled using a liquid crystal attenuator. Scale bars, 0.5 mm. Corresponding SEM images of the microstructures (i, iii, v, vii, and ix), showing the change of the morphology of the unit cell from disk to sphere and eventually a hole. Scale bars, 200 nm. Gamut loop of laser-printed structural colors, which covers CMY colors.~\cite{zhu2017resonant} (c) Temperature-feedback laser reshaping of Si resonators. SEM images show evolution of the geometric shape of the Si nanoparticle upon heating with a CW laser at intensity increasing from 4 to 122~mW/$\mu$m$^2$. Bottom: Corresponding thermal-induced shift of the c-Si Raman band upon laser heating of the isolated Si resonator. The inset shows schematic of the laser heating and Raman signal generation processes.~\cite{aouassa2017temperature}}
  \label{fig:reshaping}
\end{figure}

One of the most promising applications of dielectric nanoparticle reshaping is the data storage, color printing, and data storage, where the present memory limit due to the current density has to be surpassed for the substantial growth of data traffic and archive applications. Indeed, according to many studies, all-dielectric nanophotonic designs exhibit optical properties very sensitive to their shape.~\cite{evlyukhin2011multipole, staude2013tailoring, evlyukhin2016optical, terekhov2017multipolar}

At higher temperatures, some intermediate case can be realized, when irreversible phase transition (e.g. oxidation) causes change of nanoparticle shape and design. In work~\cite{nagasaki2018control}, lithographically fabricated Si nanoparticles were thermally oxidized in a furnace at 750 $^o$C with introducing saturated steam and air at atmospheric pressure to form an oxidized layer on the nanostructure surface. At room temperature, the Si interior is protected by the formation of a native thin oxide film (SiO$_2$), even if exposed to air for a long time.\cite{morita1990growth} However, with a sufficient amount of steam and oxygen at high temperatures, molecules diffuse through the oxide film and react with Si at the Si/SiO$_2$ interface according to the reactions Si + O$_2$ $\rightarrow$ SiO$_2$ and Si + 2H$_2$O $\rightarrow$ SiO$_2$ + 2H$_2$, resulting in an approximately 2.2-fold volume expansion of the oxide film, as shown in schematic of Fig.~\ref{fig:reshaping}a. The Si nanostructure arrays with controllable oxide layer exhibited distinct and vivid colors, which were strongly dependent on the oxidation time. Optical image in Figure~\ref{fig:reshaping}a shows wide gamut of the created Si metasurface colors, as well as the ability to the hidden pattern development after applying high temperature. Generally, compared to plasmonic analogs, color surfaces with high-index dielectrics, such as Si or Ge, have a lower reflectance, yielding a superior color contrast.~\cite{proust2016all, sun2017all, nagasaki2017all, wood2017all, vashistha2017all, flauraud2017silicon, sun2018real}

In the case of absence of considerable chemical reactions during nanostructure heating, a dewetting mechanism related to the minimization of the total energy surfaces occurs~\cite{thompson2012solid,ye2011templated,suh2002capillary}. The dewetting is a prospective tool for large-scale nanofabrication and thermally-induced reshaping of dielectric nanoparticles.~\cite{abbarchi2014wafer, naffouti2016fabrication, wood2017all, ye2019dewetting}  The laser-matter interaction responsible for the reshaping via dewetting is also related to local melting~\cite{sundaram2002inducing,bauerle2013laser}. 
This concept was successfully realized for laser-postprocessing of a Ge metasurfaces with morphology-dependent resonances and color~\cite{zhu2017resonant}, as shown in Fig.~\ref{fig:reshaping}b. Moreover, the authors used a polarization-sensitive color palette to create complex colorized images at resolutions beyond the diffraction limit. By elongating the disks into bars they showed that the asymmetric structures can support tunable color under polarized incident light in the visible spectrum. The patterns were printed with a resolution of 100 000 DPI.~\cite{zhu2017resonant} Similar concept was demonstrated with Si nanostructures.~\cite{berzinvs2020laser} Laser-induced dewetting was also applied for fabrication of metasurface spanning across 10$^2$–10$^4$ $\mu$m$^2$ and, subsequently, consisting of thousands of polycrystalline Si nanostructures distributed in an ordered rectangular lattice.~\cite{berzins2020direct}

Remarkably, that according to Section 4 on nanothermometry, the laser-induced nanoparticles reshaping can be carried out with precise temperature control by doing nanothermometry via measuring \textit{in situ} Raman signal during the process. Figure~\ref{fig:reshaping}c shows the correlation between shape of Si nanoparticle and its Raman spectrum exhibiting temperature-driven shift allowing to determine actual temperature during the reshaping.~\cite{aouassa2017temperature} Additional spatial Raman mapping of the heated nanoparticle with sub-diffractional resolution revealed stronger heating of their top parts relatively the bottom ones.


\section{Conclusion and outlook}

We have demonstrated how all-dielectric nanophotonics can be employed as a promising low-loss platform for light-matter manipulation at the nanoscale, more specifically as a powerful tool for subwavelength optical heating. The ability to tune precisely optical losses in dielectrics in a broad spectral range allows to achieve optimal conditions for light absorption in nanostructures employing the critical coupling concept when radiative and nonradiative losses become equal. Here, we have reviewed different advanced techniques recently emerged in nanothermometry dealing with all-dielectric nanoscale structures, which make such nanostructures perfect candidates for the development of 'all-in-one' platforms for simultaneous optical heating, thermometry, and additional photo-induced manipulation of various objects at the nanoscale.

We anticipate a rapid progress of the field of all-dielectric thermonanophotonics and its applications in photonics, optoelectronics, chemistry, and biomedicine. Below, we mention just a few example of the expected developments. 

First, further progress should be demonstrated in nanothermometry, where novel designs and materials for highly sensitive temperature measurements at the nanoscale (such as resonant nanodiamonds with NV-centers~\cite{kucsko2013nanometre, zalogina2018purcell} empowered by Mie resonances) can play an important role. For example, the optimization of optical tweezers for operating with various nanoscale objects would enable simultaneous temperature control~\cite{karpinski2020optical, odebo2020optical} providing a novel platform for advanced changing and probing of local temperature via all-optical manipulation with nanothermometry.

In nonlinear optics, thermo-modulated nonlinear nanophotonic designs for harmonics generation and multiphoton photoluminescence can be developed based on thermally-driven variation of the refractive index or light emission quantum efficiency. It can open new directions such as thermally tunable nonlinear flat optics for IR imaging and LIDAR applications.

In optoelectronics, smart nanostructuring of semiconductors can lead to a breakthrough in the optimization of thermo-electrical effects,~\cite{sharma2020metasurfaces} where the optimal absorption and temperature distribution at nano-/microscales are highly desirable. Moreover, thermal management is crucial for optimization of nanolasers and nano-LEDs, where overheating is a parasitic effect reducing light-emitting properties.

Strong subwavelength near-field localization and fast optical heating are highly desirable for heat-assisted data storage (e.g. magnetic recording~\cite{challener2009heat}). In this regard, advanced all-dielectric nanostructures may be useful for the precise optical heating with high spatial and temperature resolution, and they may allow for near-threshold operation of a recording device making the storage as dense as possible.

For applications dealing with functional liquids, we mention 
that a precise control on  variation of local temperature in dielectric nanoparticles can be beneficial for temperature-induced liquid flows. Besides that, the generation of temperature gradients by nanoparticles along solid/liquid interfaces can lead to thermo-osmotic flows with the speeds exceeding 100~$\mu$m s$^{-1}$, also being strongly confined to the interfacial region that is important for thermophoresis.
Stronger heating is useful for photothermally-induced microbubble formation around nanoparticles, where a bubble can nucleate at the interface above a certain power. It can also be used for various biomedical applications such as cancer therapy, photo-acoustic imaging, or optoporation.

Chemistry is also the research field that can benefit from the use of various dielectric nanoparticles. Optically resonant dielectric nanoparticles allowing for \textit{in situ} nanothermometry might find  useful applications in photothermally-assisted chemical vapour deposition, photothermally-assisted catalysis in the gas phase, as well as photothermally-assisted enhanced reactivity in solutions.

Finally, various multidisciplinary directions can emerge from the development of thermonanophotonics employing the approaches from radiophysics, where heating due to electromagnetic waves was very well known for many decades, and now it inspires nanophotonics and multidisciplinary communities to find new effects and completely unexpected applications. For example, we mention the recent studies of the formation of plasma due to electromagnetic hot-spots arising from the cooperative interaction of Mie resonances in the individual spheres, suggesting a method to experimentally model subwavelength field patterns using thermal imaging in macroscopic dielectric systems~\cite{khattak2019linking}).

\section*{Acknowledgements}

This work was supported by the Russian Science Foundation (project 21-75-30020) and the Australian Research Council (grant DP200101168). 

\section*{Appendix. Symbols and Acronyms}

\begingroup 
\renewcommand{\arraystretch}{1.2}
\begin{longtabu} to \textwidth {r X}
\textbf{NP} & Nanoparticle \\
\textbf{DF} & Dark-field scattering \\
\textbf{CW} & Continuous-wave laser \\ 
\textbf{ED, EQ, EO} & Electric dipole, quadrupole and octopole modes \\
\textbf{MD, MQ, MO} & Magnetic dipole, quadrupole and octopole modes \\
\textbf{SRS} & Stimulated Raman scattering \\
\textbf{SEM} & Scanning electron microscopy \\
\textbf{TEM} & Transmission electron microscopy \\
\textbf{AFM} & Atomic force microscopy \\
\textbf{SERS} & Surface-enhanced Raman scattering \\
{$\textbf{J}$} & Electric current \\

{$C_e,~C_i$} & Specific heat capacity of electron and lattice subsystems \\
{$T_e,~T_i$} & Temperature of electron and lattice subsystems \\
{$\gamma_{ei}$} & Coupling factor of electron-phonon interaction \\
{$N_e, N_{th}$} & Conduction band electron density and critical carrier density \\
{$G_e$} & Electron generation rate \\
{$R_e$} & Electron relaxation rate \\
{$\mu_e$} & Electron mobility \\
{$\tau_0,~\tau_{\gamma}$} & Hot carrier relaxation time and total electron-phonon relaxation time\\
{$K_c$} & Cooling rate \\
{$\rho, c_p, \kappa$} & Material density, heat capacity at constant pressure and thermal conductivity \\
{$\alpha,~\chi$} & Absorption coefficient and thermal diffusivity ($\kappa/\rho c_p$) \\
{$Q$} & and total dissipated power\\
{$R$} & Reflection coefficient \\
{$I,~I_0,~\mathfrak{P}$} & Intensity distribution inside the medium, incident light intensity and $\mathfrak{P}=a^2 I_0$ is total power of Gaussian beam with $a$ radius and $I_0$ intensity \\
{$T,~T_0,~T'$} & Temperature at specific point of the material, initial temperature and the temperature increase under laser irradiation ($T-T_0$)\\
{$C_{sca},~C_{ext},~C_{abs}$} & Scattering, extinction and absorption cross-sections of light \\
{$a_l,~b_l$} & Electric and magnetic scattering Mie-coefficients of different order\\
{$\tau_p$} & Pulse duration \\
{$\lambda$} & Wavelength of light in free space \\
{$\varepsilon,~\varepsilon',~\varepsilon''$} & Total, real part and imaginary part of dielectric permittivity \\
{$\gamma_{rad},~\gamma_{Ohmic}$} & Radiative and nonradiative (Ohmic) optical losses of the system \\
{$\textrm{n}$} & Refractive index \\

\end{longtabu} 
\endgroup

\newpage

\bibliographystyle{unsrt}
\bibliography{references}

\end{document}